\documentclass[10pt,journal,final,twoside]{IEEEtran}

\usepackage[utf8]{inputenc}
\usepackage{xspace,empheq,fancybox,amsmath,amssymb,amsfonts,epstopdf,epsfig,syntonly,times,amsthm} 
\usepackage{psfrag,color,bm,array,tabularx, subfigure}
\usepackage{cite,url,footnote,xspace,syntonly}
\usepackage{verbatim,multirow}
\usepackage[linesnumbered,ruled,vlined]{algorithm2e}
\usepackage[T1]{fontenc}
\usepackage{textcomp}
\usepackage{xcolor}
\usepackage{mathtools}
\usepackage{makecell}
\usepackage{bbm, booktabs, makecell, subcaption, graphicx, titlesec,adjustbox,lipsum}

\usepackage{cuted}
\usepackage{nomencl}
\makenomenclature
\newcolumntype{M}[1]{>{\centering\arraybackslash}m{#1}}
\newcolumntype{N}{@{}m{0pt}@{}}

\def\BibTeX{{\rm B\kern-.05em{\sc i\kern-.025em b}\kern-.08em
    T\kern-.1667em\lower.7ex\hbox{E}\kern-.125emX}}

\usepackage{tikz}

\input{mysymbol.sty}
\allowdisplaybreaks

\usepackage{algorithmic}

\newcommand{\Pfdroop}{\text{$P$-\,$\omega$ }}
\newcommand{\QVdroop}{\text{$Q$-$V$ }}

\title{Operational Risks in Grid Integration of Large Data Center Loads: Characteristics, Stability Assessments, and Sensitivity Studies}
\author{\IEEEauthorblockN{ Kyung-Bin Kwon$^*$, Sayak Mukherjee$^*$, Veronica Adetola
}
\thanks{Authors are with the Pacific Northwest National Laboratory, Richland, WA 99352, USA, Emails: \{kyung-bin.kwon, sayak.mukherjee, veronica.adetola\}@pnnl.gov, $^*$ contributed equally.  }%
\thanks{\protect\rule{0pt}{3mm}  The research is supported by the Energy and Environment Directorate's Laboratory Directed Research at Pacific Northwest National Laboratory (PNNL). PNNL is operated by Battelle for the U.S. Department of Energy under Contract DEAC05-76RL01830. }
}

\begin{document}
\maketitle

%

\begin{abstract}
This paper investigates the dynamic interactions between large-scale data centers and the power grid, focusing on reliability challenges arising from sudden fluctuations in demand. With the rapid growth of AI-driven workloads, such fluctuations, along with fast ramp patterns, are expected to exacerbate stressed grid conditions and system instabilities. We consider a few open-source AI data center consumption profiles from the MIT supercloud datasets, along with generating a few experimental HPC job-distribution-based inference profiles. Subsequently, we develop analytical methodologies for real-time assessment of grid stability, focusing on both transient and small-signal stability assessments. Energy-flow-like metrics for nonlinear transient stability, formulated by computing localized data center bus kinetic-like flows and coupling interactions with neighboring buses over varying time windows, help provide operators with real-time assessments of the regional grid stress in the data center hubs. On the other hand, small-signal stability metrics, constructed from analytical state matrices under variable operating conditions during a fast ramping period, enable snapshot-based assessments of data center load fluctuations and provide enhanced observability into evolving grid conditions. By quantifying the stability impacts of large data center clusters, studies conducted in the modified IEEE benchmark $68-$bus model support improved operator situational awareness to capture risks in reliable integration of large data center loads.
\end{abstract}


%

\section{Introduction}

AI data centers are rapidly becoming major electricity consumers, with U.S. data centers comprising $4.4$\% of national electricity use in 2023 and projected to reach $9-12$\% by 2030 \cite{council2025assessment, aljbour2024powering}. This sharp growth, driven by the adoption of generative AI like large language models (LLMs) \cite{openai2023gpt4, touvron2023llama, awais2025foundation}, introduces frequent energy fluctuations. Classified as large dynamic digital loads (LDDLs), their massive demand and sudden load swings pose serious challenges to grid stability and reliability, thus risking national energy security.
LDDL characteristics stem from server rack operations like training and inference \cite{huang2021modeling}, creating sharp load ramps. These ramps stress grid transfer capabilities, leading to congestion and both transient and small-signal stability issues. Recognizing this, the North American Electric Reliability Corporation (NERC) Large Load Task Force recently highlighted critical reliability challenges from emerging LDDLs like AI training facilities \cite{nerc2025_large_loads}.

Modeling of data center loads has also been a focus of recent research works. A substantial body of literature examines the long-term dynamics of data centers, with a primary focus on heat transfer and thermodynamic aspects \cite{khalaj2016energy, an2022dynamic}. Such research emphasizes the internal operation of LDDLs rather than their dynamic interactions with the power grid. \cite{drovtar2024utilizing} introduces a dynamic load model for voltage and reactive power control tailored to data centers, while \cite{sun2022dynamic, suryanarayana2021system} analyze data center behavior from the converter dynamics point of view. \cite{jimenez2025data} presents a dynamic load model for transient stability assessments. While transient stability analysis has been well-established in power systems literature \cite{chang1995direct, chiang1994bcu, vu2015lyapunov, backhaus2014efficient}, these methods primarily address conventional load and generation patterns. The unique rapid load transitions characteristic of AI data centers present fundamentally different transient stability challenges that remain largely underexplored in existing research. These rapid and substantial load variations can also trigger inappropriate responses from local controllers of inverter-based resources and other generation assets, potentially leading to large-scale system instabilities. A recent study led by Dominion Energy documented a real-world case of such oscillatory behavior driven by data center loads \cite{mishra2025understanding}. \cite{biswas2025evaluating} presents the risks of forced oscillations in the Western US power grid in the presence of cyclical load consumption patterns of data centers. In addition, \cite{ko2025wide} develops a dynamic power profiling approach for AI-centric loads and analyzes their potential to induce wide-area grid oscillations.
The Accurate modeling of behaviors like Fault-Ride-Through (FRT) \cite{jimenez2025data} is also essential for developing mitigation strategies, such as storage-based smoothing \cite{kundu2025managing} and workload shifting. The high-frequency power fluctuations from AI workloads are distinct from conventional demand, requiring advanced methods to assess their grid impact.

A significant gap persists in creating a comprehensive framework for real-time grid stability assessment under LDDLs. Current research is often limited to isolated component modeling or post-event analysis. A pressing need exists for analytical methods to quantitatively evaluate stability in real-time, linking AI workload patterns to actionable reliability metrics. Such a framework would enable system operators to proactively manage grid security. This paper addresses this void by proposing real-time assessment tools to characterize and quantify the stability risks posed by LDDLs.

\textit{Contributions:} The paper aims to characterize various forms of interactions between the power grid and data centers, driven by large, sudden fluctuations in demand. Its primary objective is to quantitatively assess reliability concerns, such as stressed grid conditions and system instabilities, that are likely to become more significant with the growing integration of AI-driven data centers. 

\begin{itemize}
    \item First, we consider generating critically stressed grid conditions by integrating large-scale data center clusters with the IEEE Benchmark 68-bus model that incorporates both power electronics-based and synchronous generation resources. Realistic load fluctuations, reflecting AI training and inference workloads, have been simulated to stress the system and trigger potential instability events. We provide a description of the grid integration modeling and generation of high-performance computing-based AI inference profiles, along with considering open source data profiles such as MIT supercloud datasets \cite{samsi2021supercloud}.

    \item Subsequently, we focus on the development of the analytical methodologies that consider the development of real-time assessment metrics, such as energy-flow-based metrics for nonlinear transient stability. We formulate the energy flow-based methodology for the regional data center hub by computing localized LDDL bus' kinetic-like energy flow, along with computing coupling flows with the neighbor buses. These metrics are computed over varying time windows and can provide essential observability over current grid conditions.

    \item Small-signal stability-based metrics have been developed with snapshot-based assessments for capturing the impacts of sharp ramp increases in the data center hubs. The small-signal metrics utilize the analytical state matrix constructions over variable operating conditions during the critical changes in the LDDL consumptions. The assessment studies and methodologies help to enhance understanding of grid–data center interactions and support improvements in operational stability. By providing operators with improved situational awareness, the findings will enable stability-informed decisions for resource dispatch, while also offering data center owners actionable recommendations to strengthen reliability.
\end{itemize}

The rest of the paper is organized as follows. Section II describes the details on AI operations, availability of open source datasets, HPC job-scheduling based inference pattern generation and grid integration of the LDDLs. Subsequently, we present the real-time assessment methodologies based on both nonlinear transient and small-signal stability with high consumption of LDDLs and sharp ramps in Section III. The methods also accompany a detailed numerical simulation in that section. Concluding remarks are provided in Section IV.  

\section{LDDL Characteristics and Integration}

\subsection{AI Data Center Operations and Hardware}

AI compute nodes rely on accelerator-centric architectures, where GPUs serve as the primary engines for large-scale parallel processing. A typical high-performance AI server integrates multiple GPUs interconnected through high-bandwidth links such as NVLink, for example, configurations with 8 NVIDIA H100 GPUs provide massive throughput for training and inference workloads. Each GPU is equipped with high-bandwidth memory (HBM), ensuring rapid data access to match the compute intensity.

Thermal management is critical, as the dense integration of accelerators and CPUs generates significant heat. Advanced cooling solutions, including optimized air-flow designs and liquid-cooling systems, are deployed to maintain stable operating conditions. The compute node architecture is carefully balanced across GPUs, CPUs, memory, and storage subsystems to maximize data throughput and end-to-end performance. At scale, multiple servers are organized within racks, forming tightly coupled clusters capable of supporting demanding AI workloads ranging from multi-trillion-parameter model training to latency-sensitive inference tasks. This rack-level organization characterizes modern AI data centers, enabling both efficiency and resilience at the system scale.

\subsection{Available Open-Source LDDL Profiles}

The LDDL profiles utilized in this study are derived from the open-source MIT supercloud dataset \cite{samsi2021supercloud, li2024unseen}, which captures real-world power consumption from a heterogeneous computing cluster. This data includes the training and inference runs of various large language models (LLMs), whose workloads are known for highly variable power demands, featuring both abrupt fluctuations and gradual changes. Such dynamic behavior makes these profiles excellent for assessing power system stability.

\begin{figure*}[t]
    \centering
    \subfigure[]{\label{fig:lel_load_soumya}\includegraphics[width=0.329\linewidth]{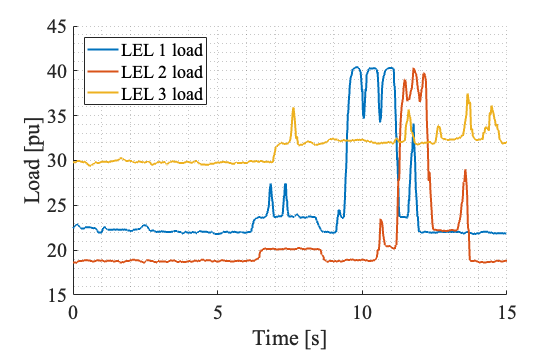}}
    \subfigure[]{\label{fig:lel_load_sai}\includegraphics[width=0.329\linewidth]{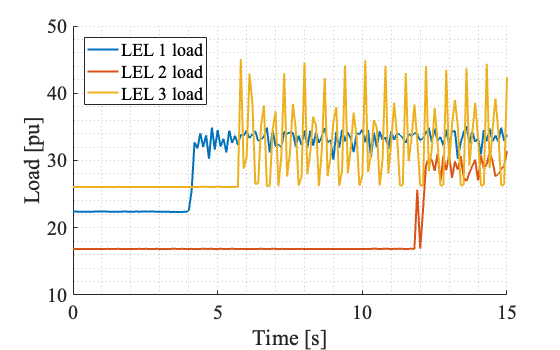}}
    \subfigure[]{\label{fig:lel_load_sayak}\includegraphics[width=0.329\linewidth]{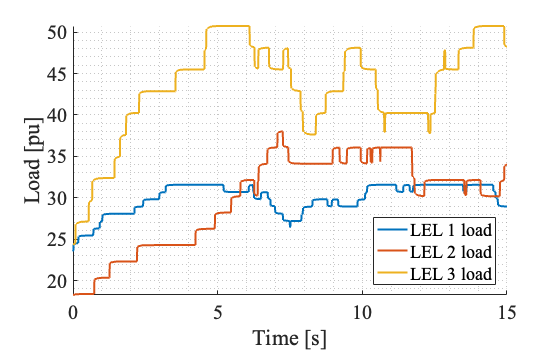}}
    \caption{Load profiles for three LELs across three distinct operational datasets: (a) Dataset A, featuring abrupt power events typical of inference tasks; (b) Dataset B, showing sustained high consumption and oscillations; and (c) Dataset C, characterized by gradual, stair-like load increases from scheduled jobs.}
    \label{fig:lel_load}
    \vspace{-0.3cm}
\end{figure*}

From this source, we curate three distinct operational scenarios, depicted in Fig.~\ref{fig:lel_load}, to model the behavior of three different LDDLs. Each dataset introduces unique disturbances to the system:

\begin{itemize}
    \item \textbf{Dataset A \cite{li2024unseen, samsi2021supercloud, kundu2025managing} (Figs.~\ref{fig:lel_load}(a)--\ref{fig:lel_load}(c)):} This dataset models short-term, high-intensity inference tasks. The profiles are characterized by sharp, high-magnitude power events. LDDL~1 and LDDL~2 experience abrupt spikes reaching peaks significantly higher than the nominal steady condition. LDDL~3 exhibits a smaller step increase with minor subsequent fluctuations.

    \item \textbf{Dataset B \cite{samsi2021supercloud} (Figs.~\ref{fig:lel_load}(d)--\ref{fig:lel_load}(f)):} This scenario represents a transition to a sustained high-consumption phase, such as the start of a model training session. LDDL~1 and LDDL~2 show sharp step increases to noisy plateaus. In contrast, LDDL~3 enters a state of persistent, high-frequency oscillations with a peak-to-peak amplitude of nearly 20\%.
    \end{itemize}

These diverse and realistic load profiles introduce substantial disturbances, providing a robust framework for assessing the impact of LDDL demand fluctuations on system dynamics. A consistent observation across all datasets is that LDDL~1 and LDDL~2 tend to exhibit greater load volatility compared to LDDL~3.

\subsection{HPC Job Scheduling-based Experimental Inference Profiles}

We consider Poisson-like HPC job arrivals for the inference tasks in an AI data center \cite{zhou2023automated, herbein2016scalable, zhang2024cross, li2022mrsch}. Let us consider the data center, where there are $M$ servers per rack with $N$ racks. $P_{base}$ denotes the idle power in kW for each of the servers, and $P_{peak}$ denotes the kW/server at full load. Let the AI job arrival rate be $\eta$ per second, and the average duration is $\gamma$ seconds. Considering the time step of $\Delta t$, the probability of job arrival in the small interval of $\eta \Delta t$ is computed, and then compared with a random number generator. Subsequently, if the AI workload has arrived, then the jobs are assigned to the idle server, which will then consume the full load. The duration of the job can follow a Gaussian distribution with mean $\gamma$, and standard deviation of $\gamma_1$ seconds. 
Therefore, for the active servers we will have for the $i-$th GPU unit in the $j-$th rack as,
\begin{align}
    & \hspace{-0.3cm} P_{GPU_{i,j}}(t) = P_{idle_{i,j}} +  P_{peak_{i,j}}, t = t_{init},1,..,\mathcal{N}(\gamma, \gamma_1),\\
    & \hspace{-0.3cm} P_{rack_j}(t) = \sum_i P_{GPU_{i,j}}(t), \; P_{AI}(t) = \sum_j P_{rack_j}(t)
\end{align}
Let us assume the desired cooling power $P_{cool}$ is set to be the $\alpha_1 P_{AI}$, then we can also write the dynamics for cooling power as:
\begin{align}
    P_{cool}(t+1) = P_{cool}(t) + \alpha_2 (\alpha_1 P_{AI}(t) - P_{cool}(t)),
\end{align}
where $\alpha_2$ is the coefficient capturing the speed of the cooling response, and $\alpha_1$ is the cooling ratio (e.g., 0.15). 
The total LDDL power is then computed as:
\begin{align}
    P_{LDDL}(t) = P_{AI}(t) + P_{cool}(t)
\end{align}
Alg. \ref{alg:hpc} shows the process of generating the inference profiles of AI data centers based on HPC job distribution routines for inference tasks. This gives us another dataset for our grid reliability experiments.

\begin{itemize}
    \item \textbf{Dataset C (Figs.~\ref{fig:lel_load}(g)--\ref{fig:lel_load}(i)):} This dataset simulates a gradual, ramp-up in load, characteristic of scheduled high-performance computing (HPC) jobs. All three LDDLs exhibit a stair-like increase, with their loads incrementally rising to their peak values from their nominal steady-state.
\end{itemize}

\begin{algorithm}[b!]
\caption{Data Center Inference Job Emulation}
\label{alg:hpc}
\begin{algorithmic}[1]

\STATE \textbf{Input:} $T, dt, \eta, \gamma, M, N, P_{peak}, P_{idle}, \alpha_1, \alpha_2$
\STATE \textbf{Initialize:} $server\_states \gets 0$, $P_{cool} \gets 0$
\STATE Initialize arrays for results

\FOR{$step = 0$ to $T-1$}
    \STATE $t \gets step \cdot dt$

    \IF{$\eta \cdot dt > random(0,1)$}
        \STATE Find idle servers
        \IF{idle servers exist}
            \STATE Assign job with duration $\mathcal{N}(\gamma, \gamma_1)$ to random idle server
        \ENDIF
    \ENDIF

    \STATE $P_{AI} \gets 0$
    \FOR{each server}
        \IF{running job}
            \STATE Add $P_{peak}$ to $P_{AI}$, decrease job time by $dt$ 
        \ELSE
            \STATE Add $P_{idle}$ to $P_{AI}$
        \ENDIF
    \ENDFOR
    \STATE $P_{AI} \gets P_{AI} \cdot N$

    \STATE $P_{cool} \gets P_{cool} + \alpha \cdot (\alpha_1P_{AI} - P_{cool})$

    \STATE $P_{LDDL} \gets P_{AI} + P_{cool}$

\ENDFOR

\STATE \textbf{Output:} $P_{AI}, P_{cool}, P_{LDDL}$

\end{algorithmic}
\end{algorithm}

\subsection{Grid Integration of the LDDLs}

\begin{figure}[t]
    \centering
    \includegraphics[width=\linewidth]{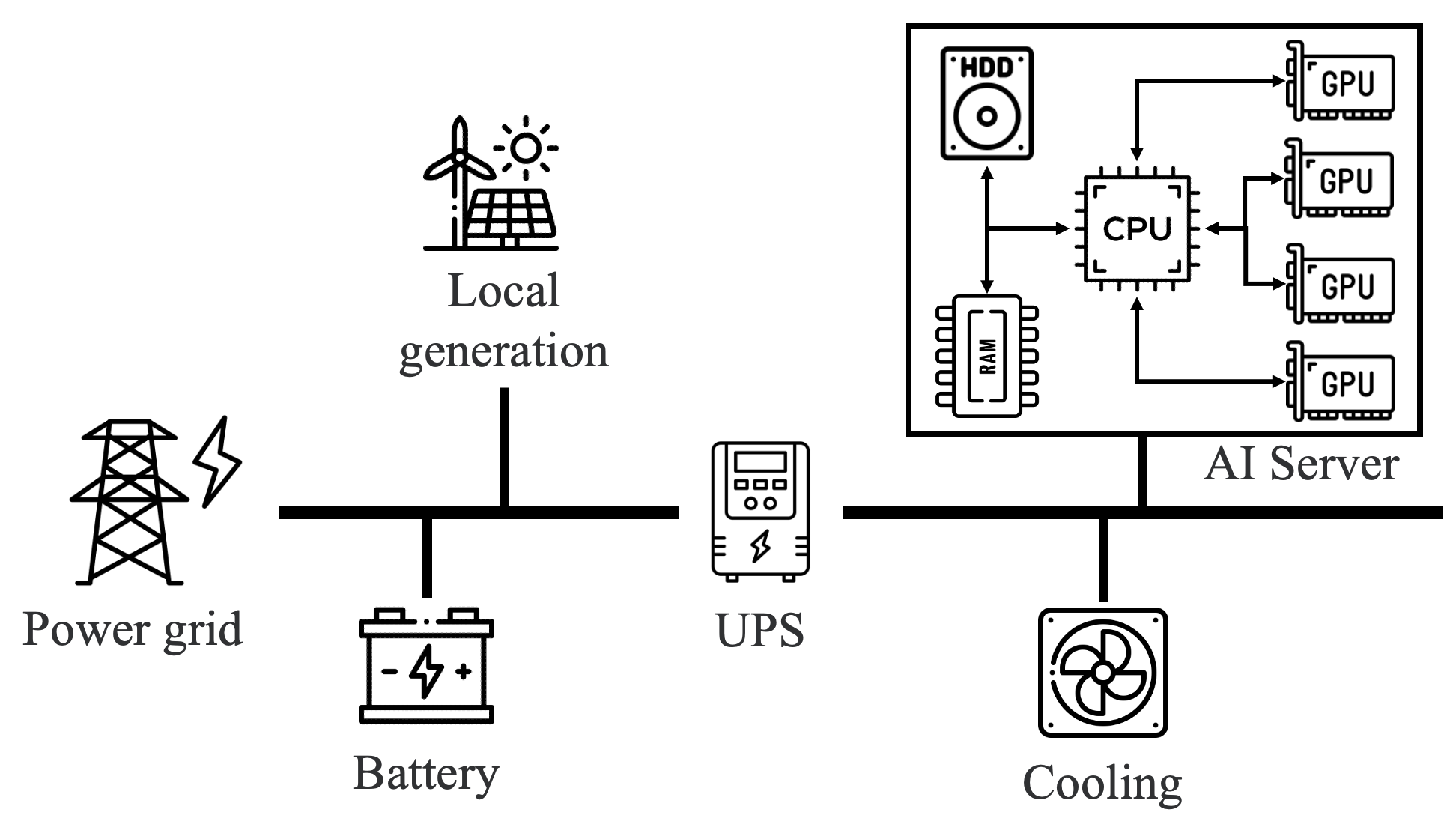}
    \caption{Grid integration overview of LDDL.}
    \label{fig:grid_integration}
\end{figure}

Data center units are integrated with the grid via a set of interconnection architectures, as depicted in Fig.~\ref{fig:grid_integration}. The data center is integrated to the utility grid via one or more high to medium-voltage feeders and step-down substation transformers. Subsequently, LDDL uses an uninterruptible power supply (UPS) that acts as a buffer between the AI workloads and the grid. Mostly, the common architecture of the UPS is to use a dual-conversion with rectifier and inverter, and the output AC is fed to the data center's dedicated power distribution units (PDUs), which then provide the power to different components within the LDDL, such as servers and cooling loads. Currently, the industry is considering utilizing grid-interactive UPS technology that can incorporate grid-forming operations to support the grid, that can provide stable voltages and frequencies, support black start capabilities, and enable integrating local generation resources such as PV, and integrating advanced coordinated storage technologies. Modeling the load characteristics of LDDL is also an active area of research, with NERC's Large Load Taskforce is discussing different modeling approximations. The dynamic nature of the LDDL consumption has been described in the previous section; apart from servers and cooling, the rest of the LDDL components can be aggregated as ZIP loads. In this study, we focus mainly on integrating various dynamic power consumption patterns in $P_{LDDL}$, and then subsequently consider an interfacing droop-controlled grid-forming dynamics mimicking provision for storage, interactive UPS, and installation of local generation. We also model the power distribution unit with a first-order filter that interfaces between the UPS and AI servers. 


\section{Analytics for Real-time Situational Awareness and Numerical Experiments}
\subsection{Nonlinear Transient Behavior} \label{sec:tran}

This section analyzes the nonlinear transient response of the power system, specifically when LDDLs are subjected to a sudden load increase. To quantify the impact of these events, we employ an energy function framework. This approach, analogous to analyzing the kinetic energy of a conventional synchronous generator, provides an insightful metric for evaluating the system's response in the presence of dynamically responsive elements like LDDLs. By deriving an energy function for the GFM inverter within an LDDL's grid interface, we can establish a clear relationship between load fluctuations and the inverter's dynamic behavior, enabling a detailed analysis of transient phenomena.

For an LDDL at bus $i \in \mathcal{L}$, the localized energy flow $E_i^l(t)$ at time $t$ is expressed as:
\begin{align}
    E_i^l(t) = \frac{1}{2} M_{eq_i} (\omega_i(t) - \omega_0(t))^2, \label{eq:local_e_flow}
\end{align}
where $M_{eq_i}$ represents the equivalent inertia of the inverter's control scheme (e.g., different forms of GFM control) and $\omega_0(t)$ is the nominal frequency. For the droop-controlled GFM inverters used in our interactive LDDL model, this equivalent inertia is calculated as:
$
    M_{eq} = 2H_{eq} = \frac{m_p}{\tau}, \label{eq:vir_inertia}
$
where $m_p$ and $\tau$ are the droop gain and the time constant of the low-pass filter in the inverter's power control loop, respectively.

In addition to this localized component, we consider the potential energy-like quantity dissipating through the transmission lines connected to the bus. This is captured by the coupling energy flow, $E_i^c(t)$:
\begin{align}
    E^c_i(t) = \frac{1}{2}\sum_{j \in \mathcal{N}_i} b_{ij} (\theta_i(t) - \theta_j(t))^2,
    \label{eq:coup_e_flow}
\end{align}
where $\mathcal{N}_i$ is the set of buses adjacent to bus $i$, and $b_{ij}$ is the susceptance of the line connecting buses $i$ and $j$. By combining these two components, we define a representative energy-like function that captures the accumulated stress over a time window $\mathcal{T}_t = [t, t+\Delta t_w]$:
\begin{align}
    E_i(t) = \sum_{t' \in \mathcal{T}_t}  (E^l_i(t') + w \cdot E^c_i(t')),
\end{align}
where $w$ is a weighting coefficient that balances the relative contributions of the local and coupling energy terms. We use the nominal load consumption amount to scale the coupling energy flow.

The energy definitions in \eqref{eq:local_e_flow} and \eqref{eq:coup_e_flow}, following from the system stability theory, are always non-negative. For practical indicator design, as they quantify the magnitude of a deviation but not its direction, we slightly modify these definitions to formulate a \textit{directional energy-like} function. The directional local energy-like flow, $E_i^{ld}(t)$, is defined as:
\begin{align}
    E_i^{ld}(t) = \frac{1}{2} M_{eq_i} \cdot |\omega_i(t)-\omega_0(t)| \cdot (\omega_i(t)-\omega_0(t)),
    \label{eq:dir_local_e_flow}
\end{align}
and similarly, the directional coupling energy-like flow from bus $i$, $E_i^{cd}(t)$, is given by:
\begin{align}
    E_i^{cd}(t) = \frac{1}{2} \sum_{j \in \mathcal{N}_i} b_{ij} \cdot |\theta_i(t) -  \theta_j(t)| \cdot (\theta_i(t) - \theta_j(t)).
    \label{eq:dir_coup_e_flow}
\end{align}
\begin{algorithm}[t!]
\caption{Energy Flow Based Analytics for Transient Stability Analysis}
\label{alg:energy_flow}
\begin{algorithmic}[1]
\STATE \textbf{Input:} Set of LDDL buses $\mathcal{L}$, adjacency sets $\mathcal{N}_i$, line susceptances $b_{ij}$, equivalent inertias $M_{eq_i}$.
\STATE \textbf{Input:} Weighting coefficient $w$, time window duration $\Delta t_w$.
\STATE \textbf{Input:} Time-series data: $\omega_i(t)$ and $\theta_i(t)$ for $t \in [0, T]$ and relevant buses.
\STATE \textbf{Initialize:} Arrays for $E^l_i(t)$, $E^c_i(t)$, $E^{ld}_i(t)$, $E^{cd}_i(t)$, $E_i(t)$, and $E^d_i(t)$.

\FOR{$t$ in time steps from $0$ to $T$}
    \FOR{each bus $i \in \mathcal{L}$}
        \STATE $\Delta \omega_{i}(t) \gets \omega_i(t) - \omega_0(t)$
        \STATE $E_i^l(t) \gets \frac{1}{2} M_{eq_i} \Delta \omega_{i}(t)^2$
        \STATE $E_i^{ld}(t) \gets \frac{1}{2} M_{eq_i} |\Delta \omega_i(t)| \Delta \omega_{i}(t)$
        
        \STATE $E_i^c(t) \gets 0$; \quad $E_i^{cd}(t) \gets 0$
        \FOR{each neighbor $j \in \mathcal{N}_i$}
            \STATE $\Delta\theta_{ij}(t) \gets \theta_i(t) - \theta_j(t)$
            \STATE $E_i^c(t) \gets E_i^c(t) + \frac{1}{2} b_{ij} (\Delta\theta_{ij}(t))^2$
            \STATE $E_i^{cd}(t) \gets E_i^{cd}(t) + \frac{1}{2} b_{ij} |\Delta\theta_{ij}(t)| \Delta\theta_{ij}(t)$
        \ENDFOR
        
         $\mathcal{T}_t = [t - \Delta t_w, t]$
        \IF{current time $t \ge \Delta t_w$}
            \STATE $E_i(t) \gets \sum_{t' = t - \Delta t_w}^{t} (E^l_i(t') + w \cdot E^c_i(t'))$
            \STATE $E_i^d(t) \gets \sum_{t' = t - \Delta t_w}^{t} (E^{ld}_i(t') + w \cdot E^{cd}_i(t'))$
        \ELSE
            \STATE $E_i(t) \gets 0$; \quad $E_i^d(t) \gets 0$ 
        \ENDIF
    \ENDFOR
\ENDFOR
\STATE \textbf{Output:} $E^l, E^c, E^{ld}, E^{cd}, E, E^d$.
\end{algorithmic}
\end{algorithm}
Lastly, we can define the total directional energy function for a time window $\mathcal{T}_t = [t, t+\Delta t_w]$ as
\begin{align}
    E^d_i(t) = \sum_{t' \in \mathcal{T}_t}  (E^{ld}_i(t') + w \cdot E^{cd}_i(t')),
\end{align}
\begin{figure}[t]
    \centering
    \includegraphics[width=\linewidth]{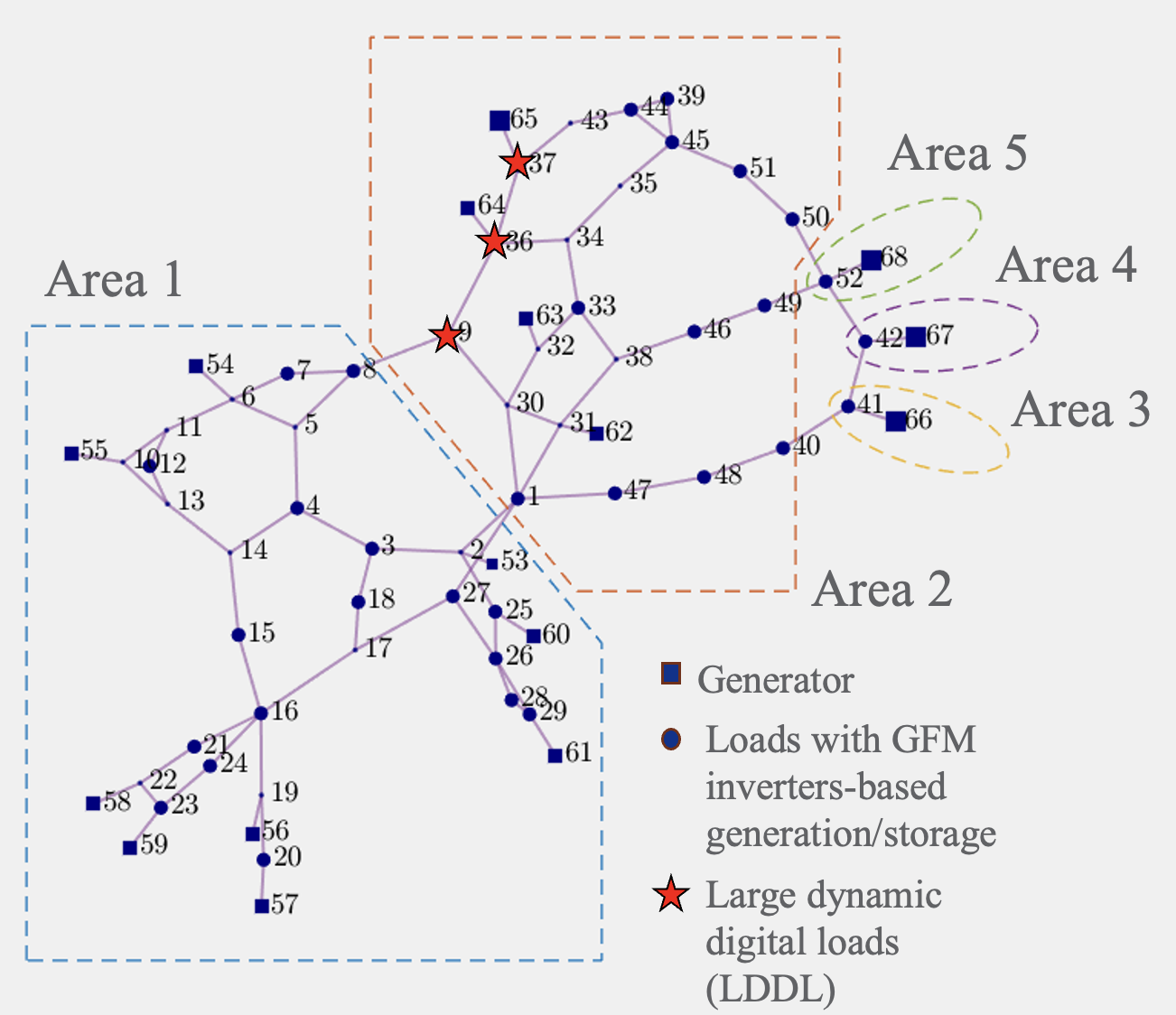}
    \caption{IEEE 68-bus system with an LDDL cluster.}
    \label{fig:system}
\end{figure}
%
%
\begin{figure*}[t]
    \centering
    \subfigure[]{\label{fig:sys_freq_data_A}\includegraphics[width=0.32\linewidth]{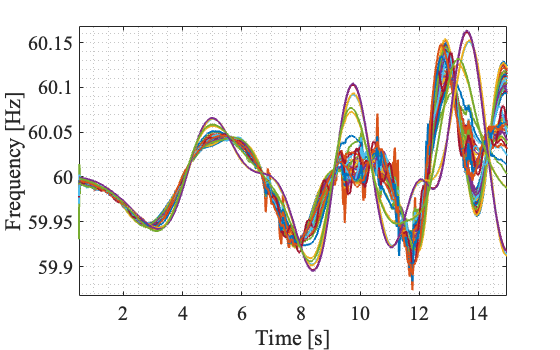}}
    \subfigure[]{\label{fig:lel_active_p_data_A}\includegraphics[width=0.32\linewidth]{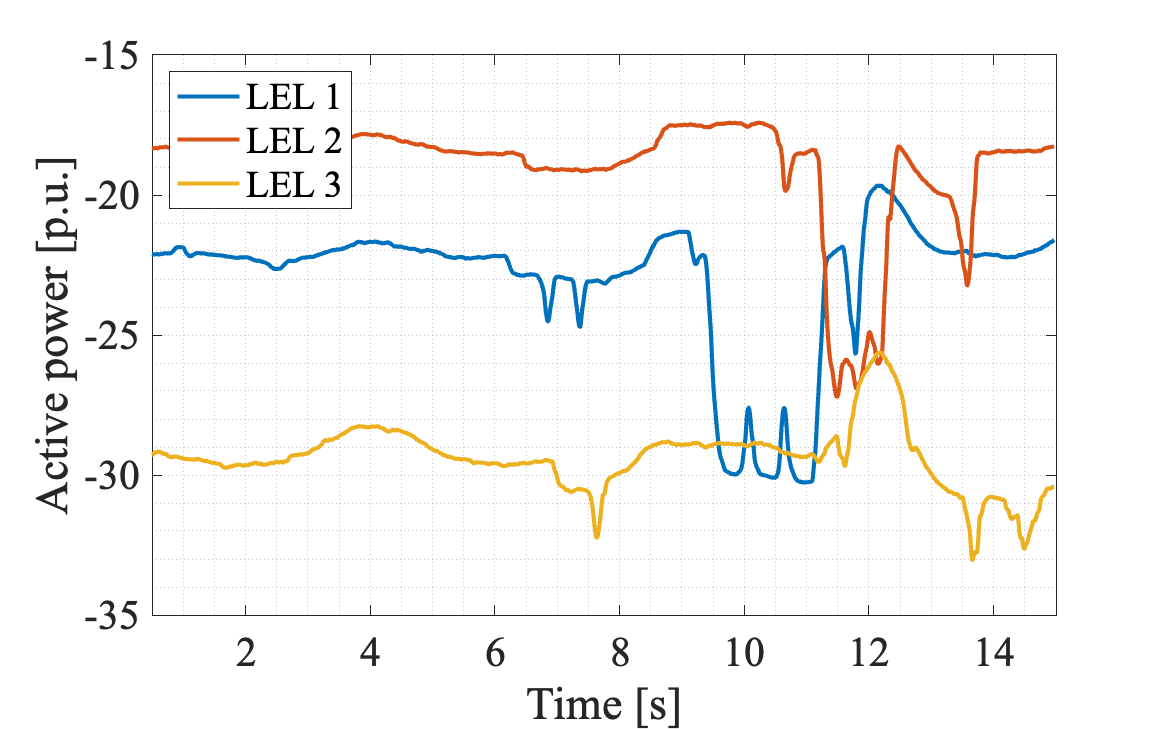}}
    \subfigure[]{\label{fig:lel_reactive_p_data_A}\includegraphics[width=0.32\linewidth]{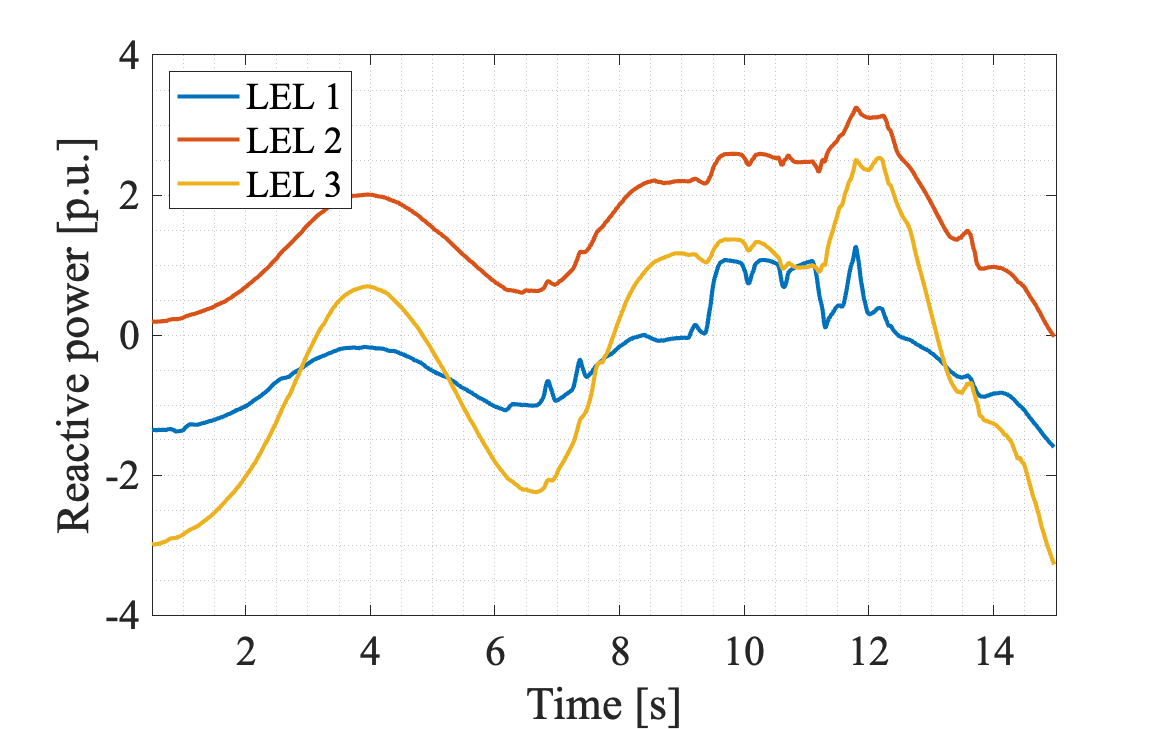}}
    \subfigure[]{\label{fig:lel_dir_energy_flow_data_A_zoom}\includegraphics[width=0.32\linewidth]{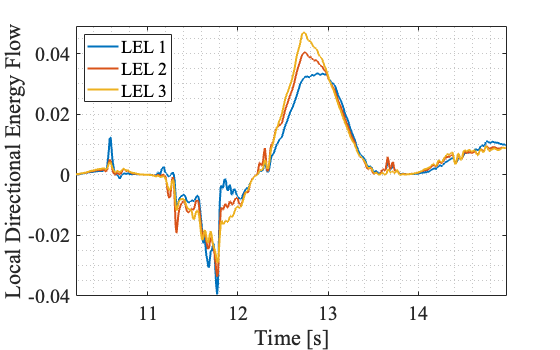}}
    \subfigure[]{\label{fig:sum_dir_energy_flow_data_A}\includegraphics[width=0.32\linewidth]{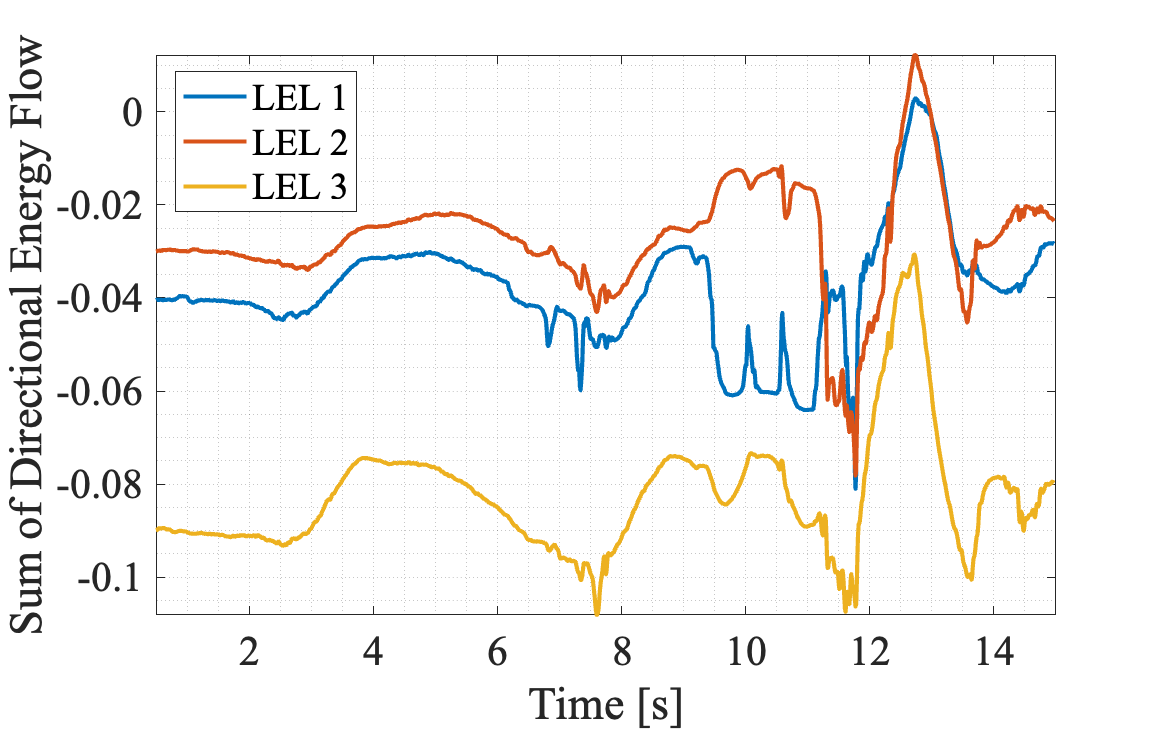}}
    \subfigure[]{\label{fig:def_bar_data_A}\includegraphics[width=0.32\linewidth]{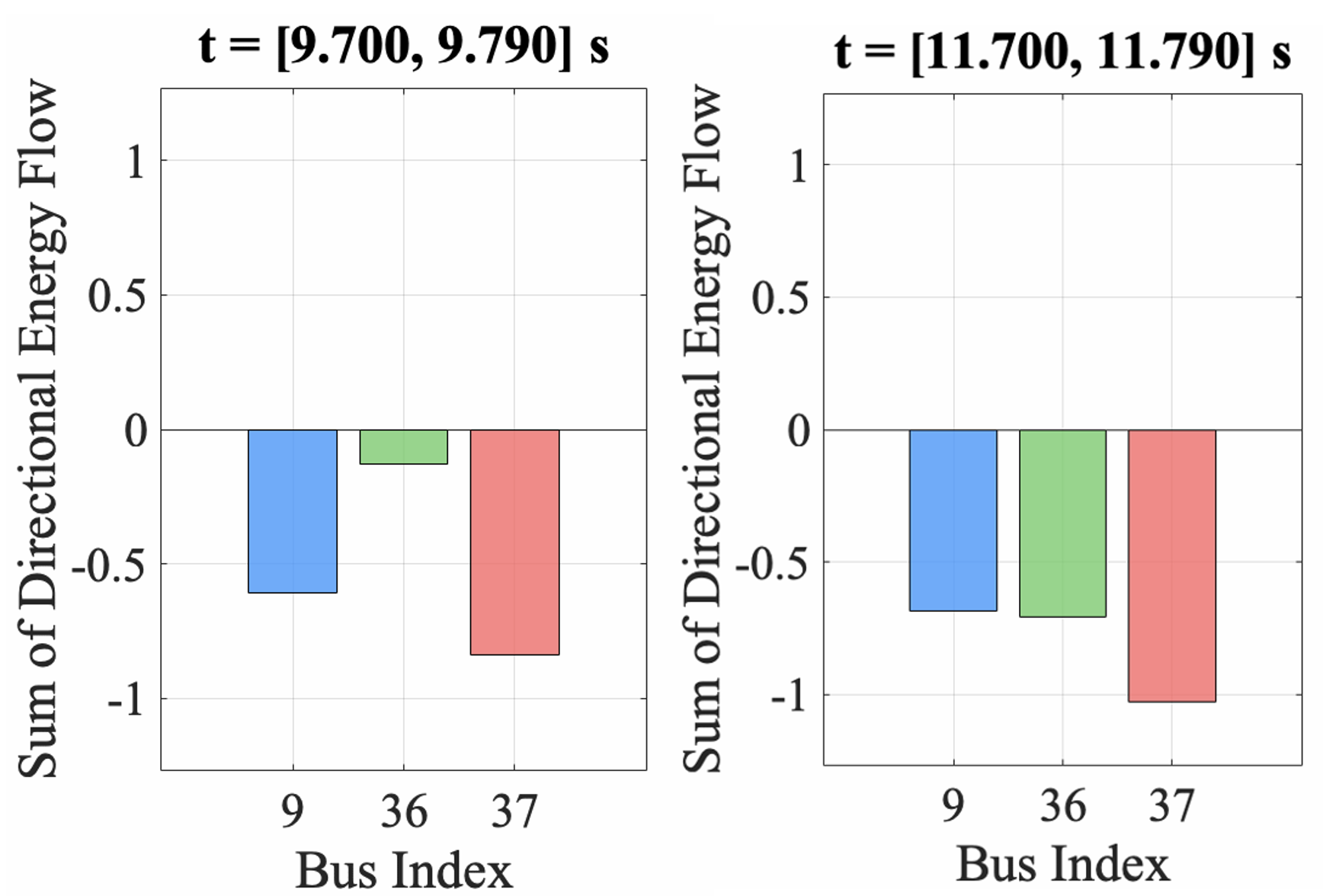}}
    \caption{Dataset A simulation result: (a) System frequency, (b) LDDL bus active power, (c) LDDL bus reactive power, (d) LDDL bus local directional energy flow, (e) total directional energy flow trajectory, and (f) Snapshot of total directional energy flow.}
    \label{fig:lel_result_data_A}
    \vspace{-0.3cm}
\end{figure*}
%
\begin{figure*}[t]
    \centering
    \subfigure[]{\label{fig:sys_freq_data_B}\includegraphics[width=0.32\linewidth]{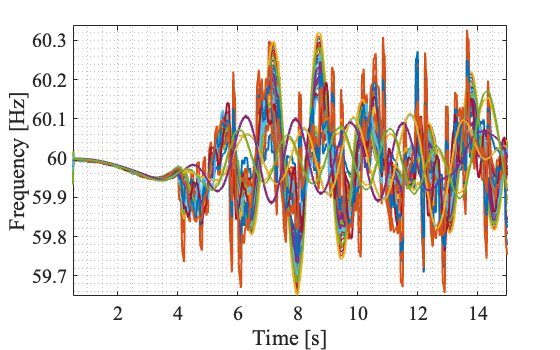}}
    \subfigure[]{\label{fig:lel_active_p_data_B}\includegraphics[width=0.32\linewidth]{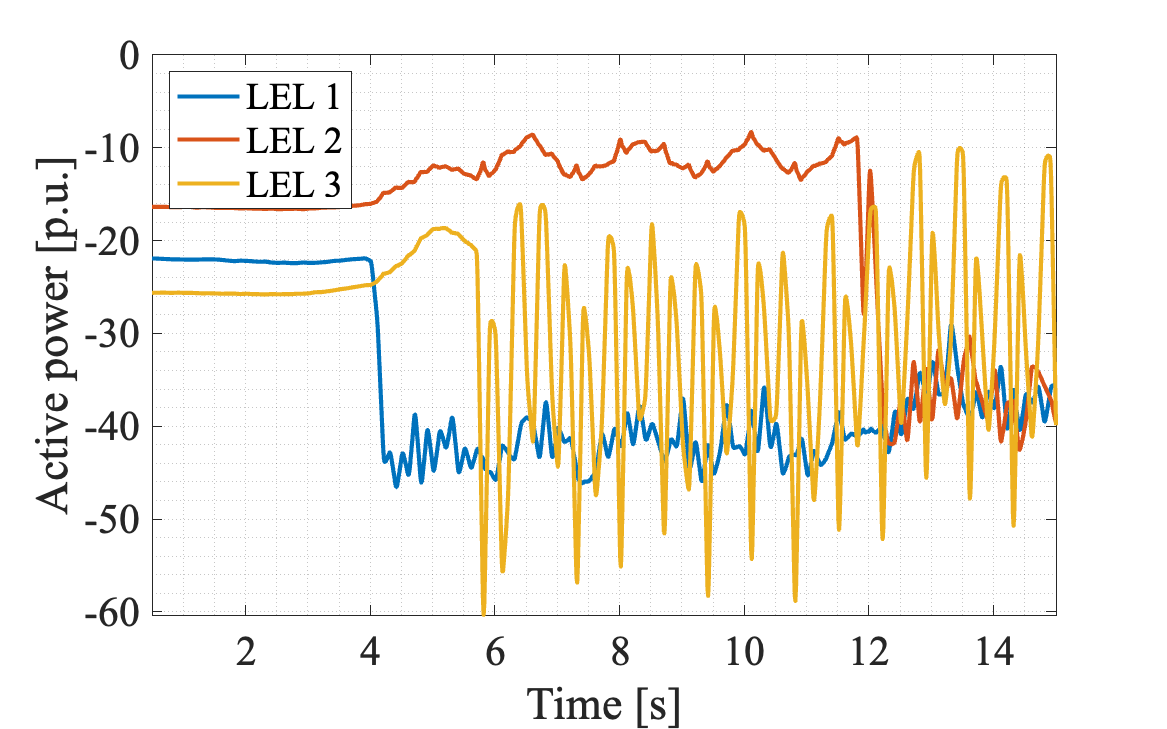}}
    \subfigure[]{\label{fig:lel_reactive_p_data_B}\includegraphics[width=0.32\linewidth]{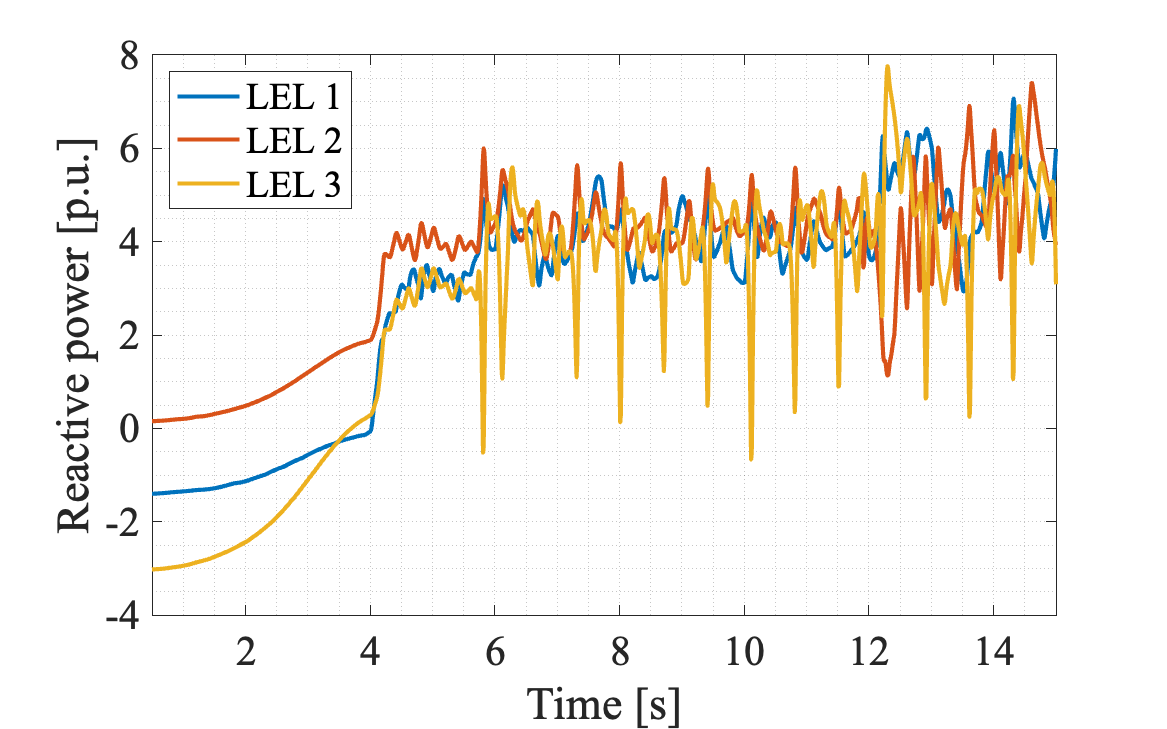}}
    \subfigure[]{\label{fig:lel_dir_energy_flow_data_B_zoom}\includegraphics[width=0.32\linewidth]{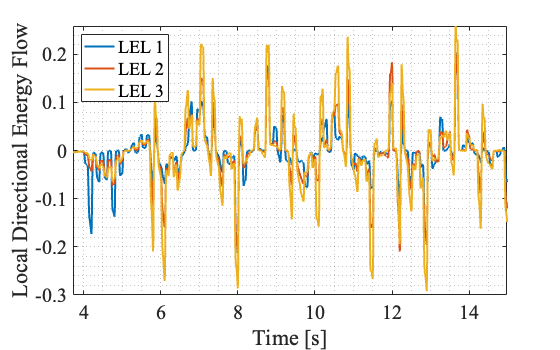}}
    \subfigure[]{\label{fig:sum_dir_energy_flow_data_B}\includegraphics[width=0.32\linewidth]{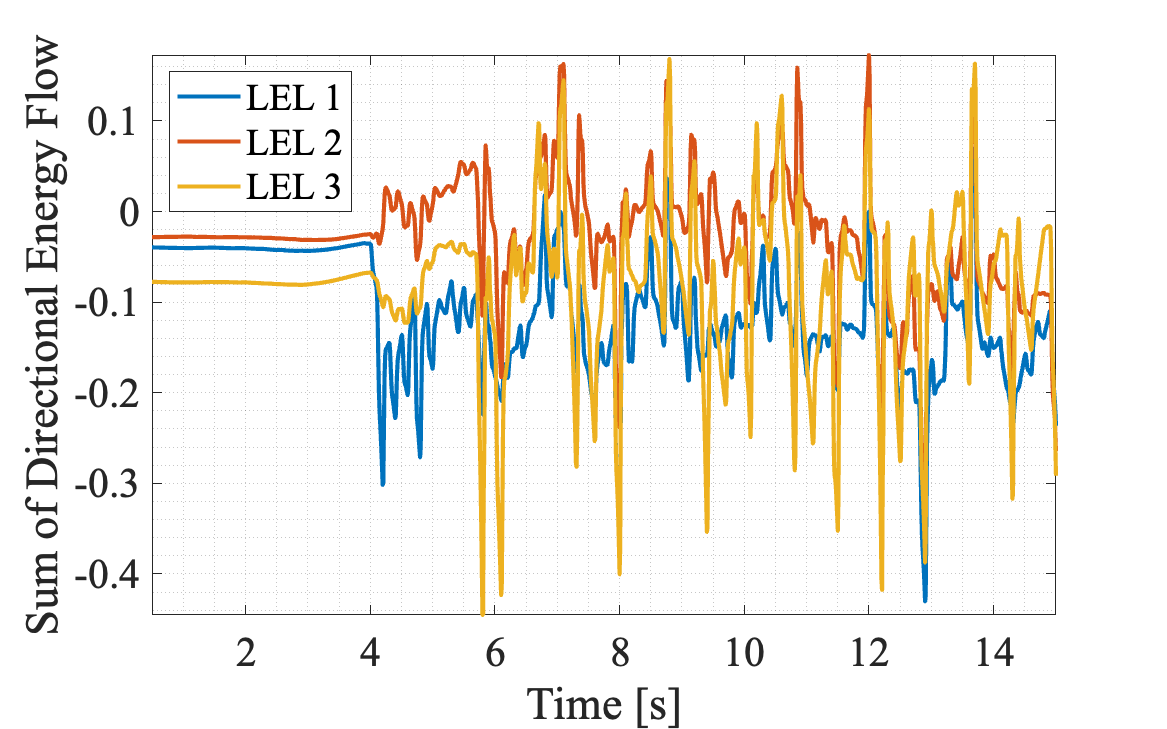}}
    \subfigure[]{\label{fig:def_bar_data_B}\includegraphics[width=0.32\linewidth]{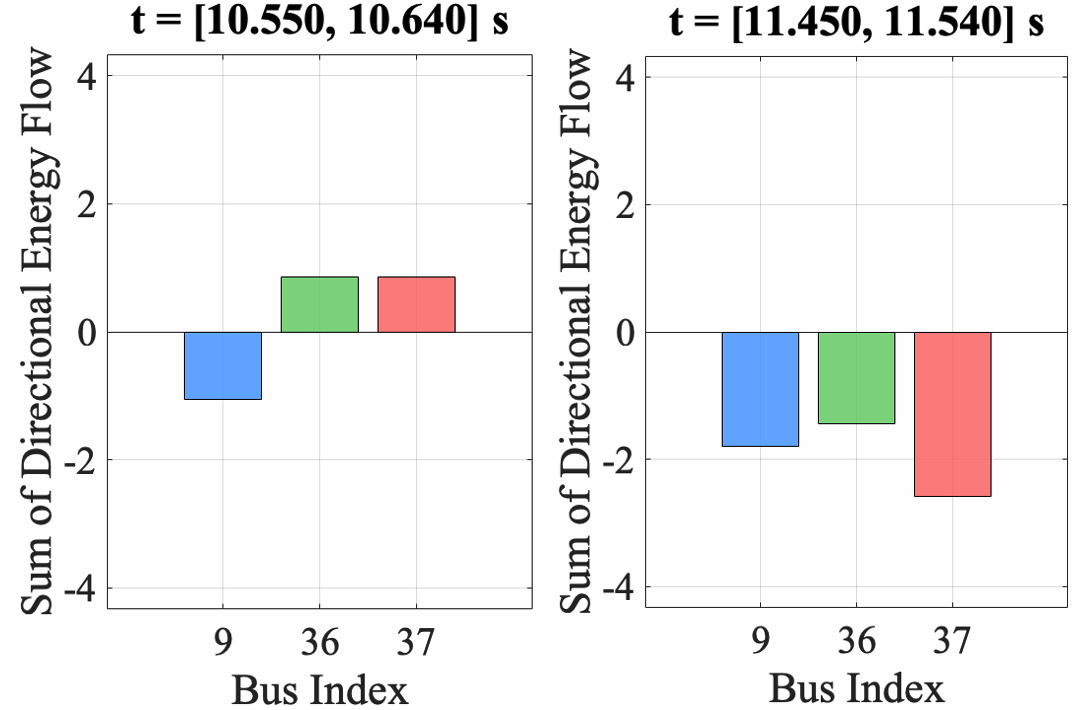}}
    \subfigure[]{\label{fig:sys_freq_data_Bcollapse}\includegraphics[width=0.32\linewidth]{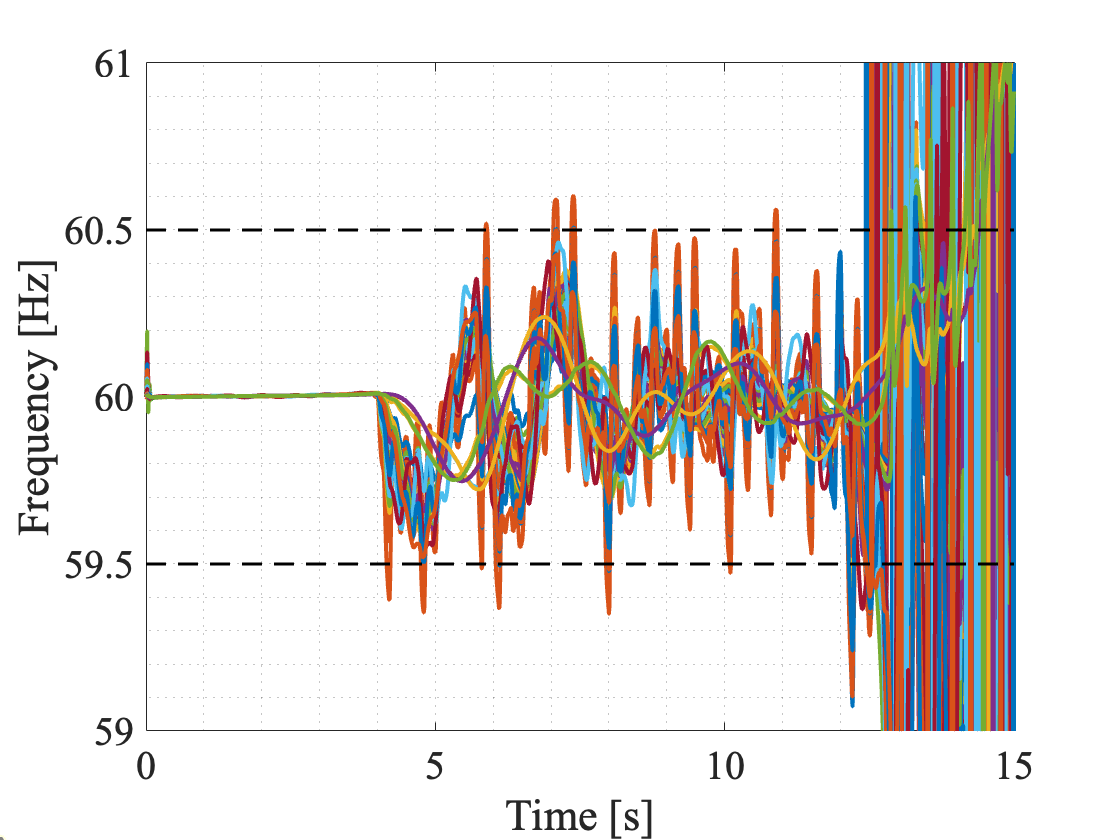}}    
    \subfigure[]{\label{fig:sum_dir_energy_flow_data_B_collapse}\includegraphics[width=0.32\linewidth]{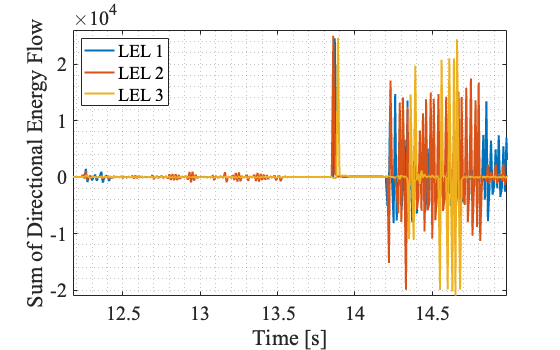}}
    \subfigure[]{\label{fig:def_bar_data_B_collapse}\includegraphics[width=0.32\linewidth]{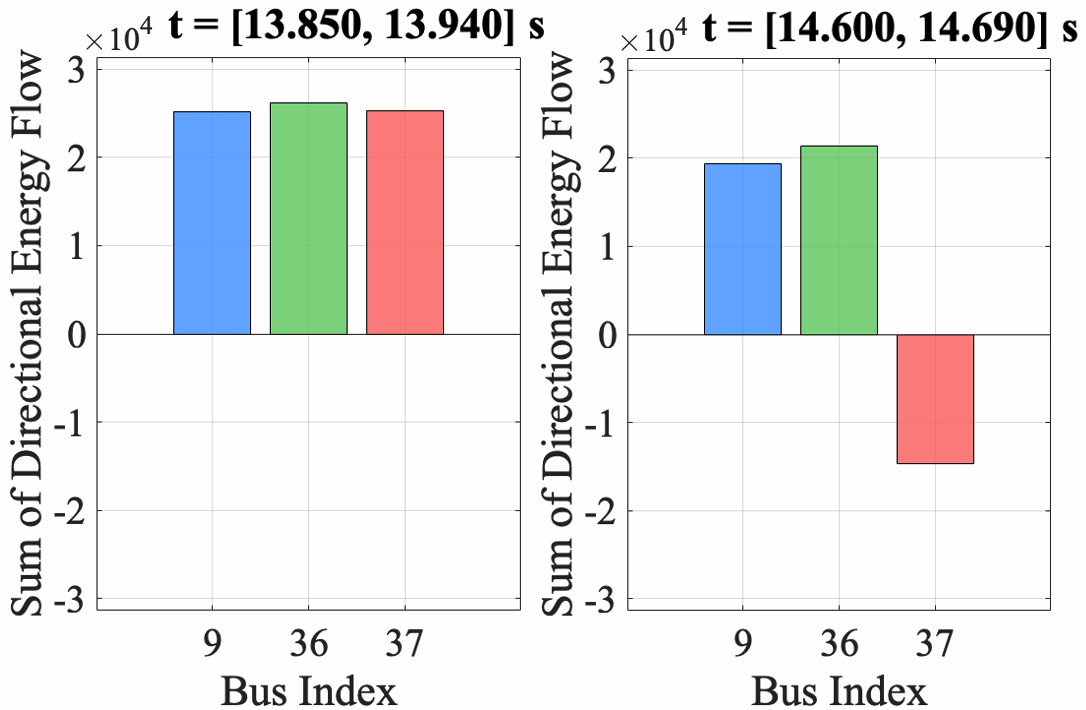}}
    \caption{Dataset B simulation result: (a)-(f) System response under standard load. (g)-(i) System collapse scenario with $1.6$x load fluctuation. Subplots show (a) System frequency, (b) Active power, (c) Reactive power, (d) Local directional energy flow, (e) total directional energy flow trajectory, (f) Snapshot of total directional energy flow, (g) Frequency collapse, (h) total directional energy flow trajectory during collapse, and (i) Snapshot during collapse.}
    \label{fig:lel_result_data_B}
    \vspace{-0.3cm}
\end{figure*}
%
\begin{figure*}[t]
    \centering
    \subfigure[]{\label{fig:sys_freq_data_C}\includegraphics[width=0.32\linewidth]{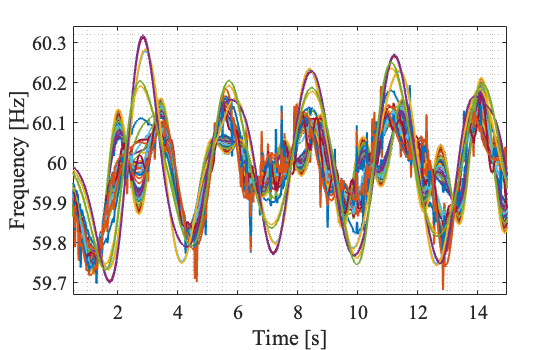}}
    \subfigure[]{\label{fig:lel_active_p_data_C}\includegraphics[width=0.32\linewidth]{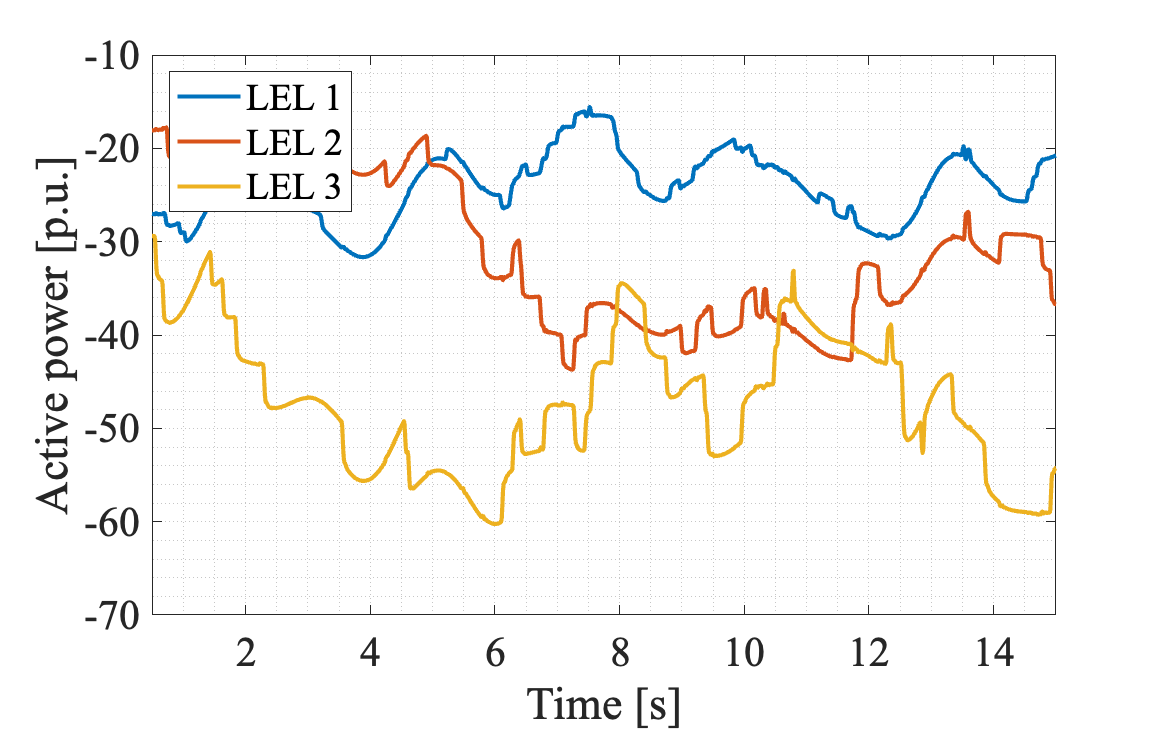}}
    \subfigure[]{\label{fig:lel_reactive_p_data_C}\includegraphics[width=0.32\linewidth]{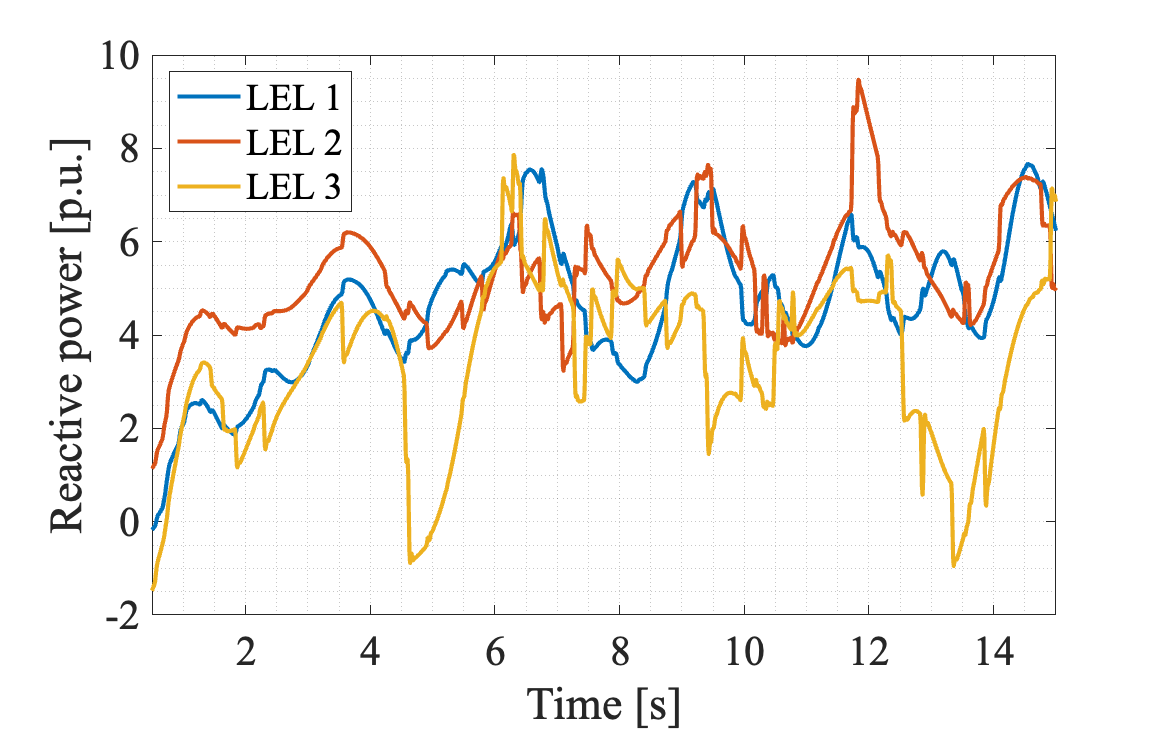}}
    \subfigure[]{\label{fig:lel_dir_energy_flow_data_C_zoom}\includegraphics[width=0.32\linewidth]{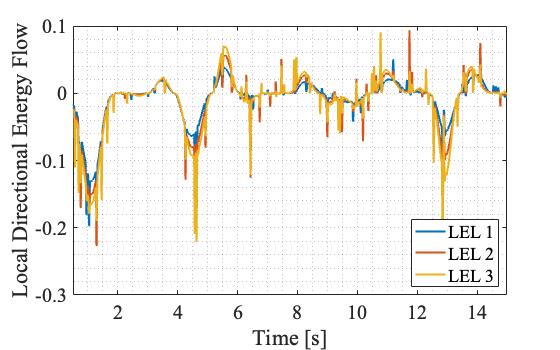}}
    \subfigure[]{\label{fig:sum_dir_energy_flow_data_C}\includegraphics[width=0.32\linewidth]{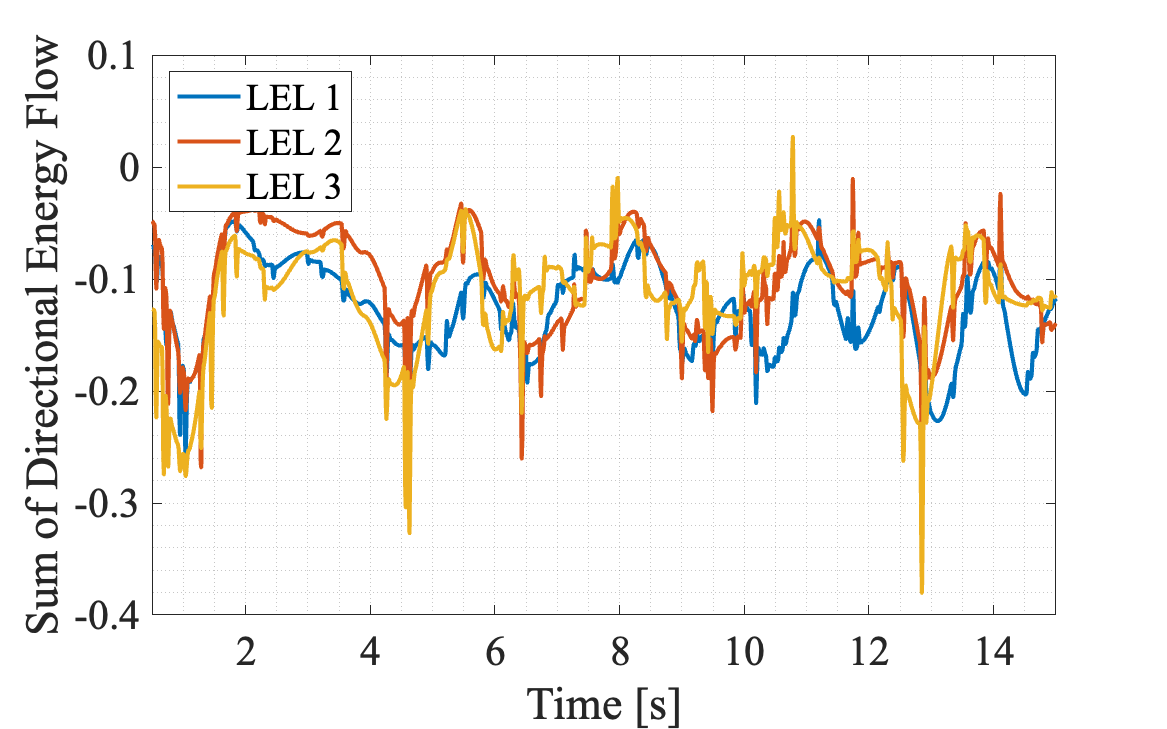}}
    \subfigure[]{\label{fig:def_bar_data_C}\includegraphics[width=0.32\linewidth]{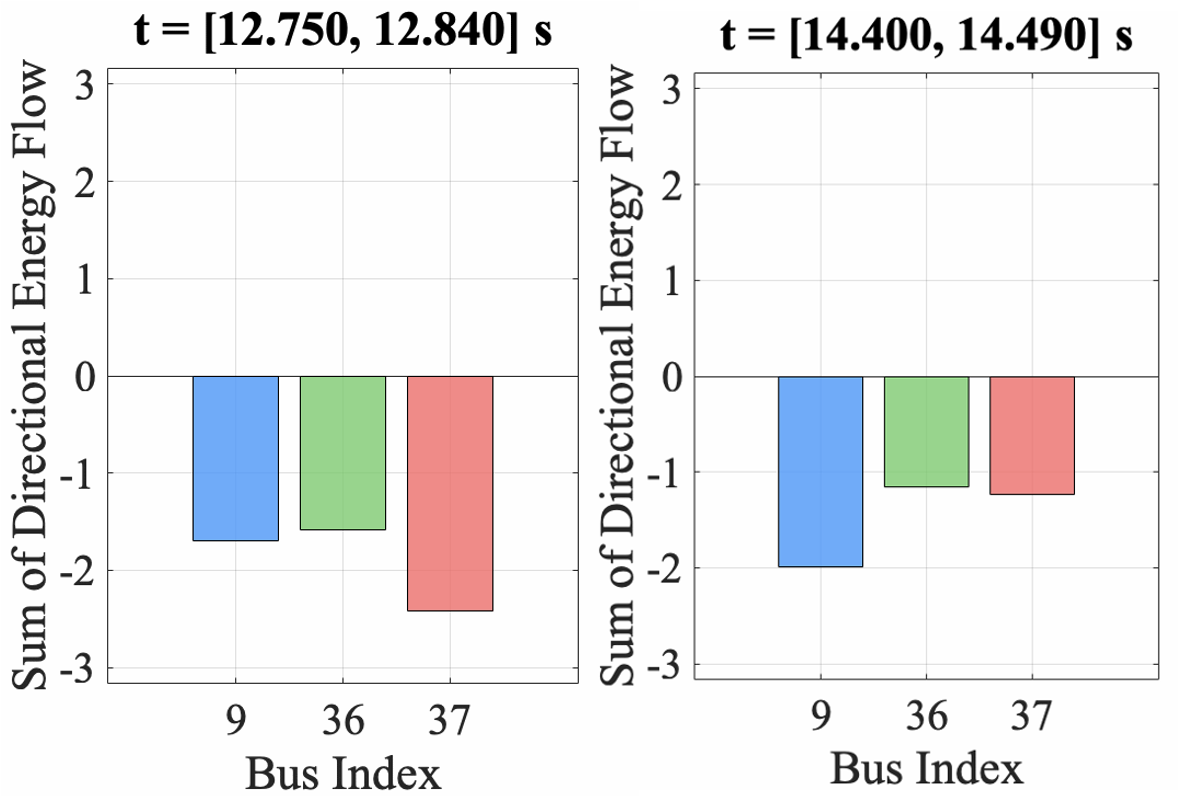}}
    \caption{Dataset C simulation result: (a) System frequency, (b) LDDL bus active power, (c) LDDL bus reactive power, (d) LDDL bus local directional energy flow, (e) Total directional energy flow trajectory, and (f) Snapshot of total directional energy flow.}
    \label{fig:lel_result_data_C}
    \vspace{-0.3cm}
\end{figure*}
These directional metrics provide not only the magnitude of the energy flow but also its direction, indicating whether energy is being absorbed by or injected into the bus and its connecting lines. For instance, a positive directional local energy ($E_i^{ld}(t) > 0$) signifies that $(\omega_i(t)-\omega_0(t)) > 0$. This implies the inverter at bus $i$ is operating at a frequency above nominal, behaving as if it possesses excess kinetic-like energy that it will tend to release back into the system. Conversely, a negative value ($E_i^{ld}(t) < 0$) indicates a frequency deficit ($(\omega_i(t)-\omega_0(t)) < 0$), where the LDDL bus is in a deficit phase of its oscillation and is absorbing energy from the grid. The interpretation of the coupling energy is analogous. A positive term in the summation for $E_i^{cd}(t)$, corresponding to $\theta_i > \theta_j$, indicates that bus $i$ is pushing active power towards the neighboring bus $j$, while a negative term implies power flow in the opposite direction.

To operationalize this comprehensive framework, we detail the procedure for calculating all defined energy metrics from system measurement data. Algorithm~\ref{alg:energy_flow} outlines the computational steps for quantifying both instantaneous and accumulated transient stress on each LDDL bus. The outputs of this algorithm serve as the primary analytics for our numerical experiments.

\subsection{Experiments with Transient Simulations}

To analyze the impact of LDDL loads on the power system, we design a test system based on the IEEE 68-bus system depicted in Fig.~\ref{fig:system} \cite{ieee68} in phasor domain. The system is modified with inverter and LDDL installations and consists of 16 generators and 35 loads with local GFM-inverter-based generation/storage, with components in a $100$ MVA base. 
Here, an LDDL cluster is established in Area 2, distributed across nodes 9, 36, and 37, referred to as LDDL~1, LDDL~2, and LDDL~3, respectively. The total LDDL loads are 35.5\%, 35.2\%, and 37.1\% of the total system load for the three test cases, respectively.
The dynamic responses of the power system to the three LDDL load datasets (from Fig.~\ref{fig:lel_load}) are illustrated in Fig.~\ref{fig:lel_result_data_A}, Fig.~\ref{fig:lel_result_data_B}, and Fig.~\ref{fig:lel_result_data_C}. Each dataset reveals distinct stability characteristics directly linked to the nature of its load profile.

\textit{Dataset A:} The sharp, periodic load spikes characteristic of inference tasks (Fig.~\ref{fig:lel_load}(a)-(c)) induce severe transient instability, as shown in Fig.~\ref{fig:lel_result_data_A}. The system frequency exhibits large, erratic fluctuations (Fig.~\ref{fig:sys_freq_data_A}). While the active power spikes of all three LDDLs create significant stress (Fig.~\ref{fig:lel_active_p_data_A}), the combined reactive power demands (Fig.~\ref{fig:lel_reactive_p_data_A}) further amplify system instability. This is reflected in the local directional energy flow (Fig.~\ref{fig:lel_dir_energy_flow_data_A_zoom}), which shows significant fluctuations between 10--12 seconds, corresponding to the primary peak load period. The total directional energy flows (local and coupling) in Fig.~\ref{fig:sum_dir_energy_flow_data_A} confirm this transient stress. The bar graphs in Fig.~\ref{fig:def_bar_data_A}, which are snapshots of the total directional energy flow, reveal critical dynamics: at times, all three LDDLs contribute negatively in unison, magnifying system instability, while at other moments, their opposing positive and negative flows create a partial cancellation effect.

\textit{Dataset B:} This scenario demonstrates the system's reaction to sustained and oscillatory load changes (Fig.~\ref{fig:lel_load}(d)-(f)). The system initially enters a stressed steady-state with persistent frequency fluctuations (Fig.~\ref{fig:sys_freq_data_B}) driven by the oscillatory load of LDDL~3. To investigate stability limits, we simulate a collapse scenario by amplifying the load fluctuations by $1.6$ times. As shown in Fig.~\ref{fig:sys_freq_data_Bcollapse}, the system frequency destabilizes and collapses after 12 seconds. This collapse is marked by a dramatic surge in the total directional energy flow, with magnitudes reaching several orders of magnitude higher than normal operating conditions (Fig.~\ref{fig:sum_dir_energy_flow_data_B_collapse}). The snapshot in Fig.~\ref{fig:def_bar_data_B_collapse} corroborates this, with energy flow metrics reaching extreme values that are significantly elevated compared to stable operation. This demonstrates that a system collapse drives the proposed energy flow metrics to very high values, serving as a clear indicator of catastrophic instability. Even during this event, instances of opposing flows among the LDDLs can be observed.

\textit{Dataset C:} In stark contrast to the erratic behavior in Dataset A, the gradual, stair-step load increases from Dataset C result in a visually more stable system response (Fig.~\ref{fig:lel_result_data_C}). This is because the slow-changing active and reactive power demands (Figs.~\ref{fig:lel_active_p_data_C} and \ref{fig:lel_reactive_p_data_C}) allow the grid's control mechanisms to adapt. Consequently, the system frequency (Fig.~\ref{fig:sys_freq_data_C}) exhibits well-damped, regular oscillations with a consistent periodicity of approximately 2-3 seconds, maintaining overall stability. The local directional energy flow (Fig.~\ref{fig:lel_dir_energy_flow_data_C_zoom}) also reflects this, showing smooth transitions rather than sharp spikes. However, the total directional energy flows (Fig.~\ref{fig:sum_dir_energy_flow_data_C}) reveal a different dynamic: it displays significant, low-frequency oscillations with wide deviation ranges that follow the system's natural oscillatory modes. The snapshots in Fig.~\ref{fig:def_bar_data_C} confirm this behavior, showing the LDDLs' energy flows oscillating in unison at a slower pace with relatively consistent amplitudes.

This comparison provides a crucial insight: the \textit{rate of change} of the LDDL load determines the \textit{type} of system stress, acting as a disturbance in the transient simulation. Rapid load variations (Dataset A) trigger high-frequency transient instability with erratic frequency deviations, sustained oscillatory loads (Dataset~B) can lead to system collapse under extreme conditions, while slow ramps (Dataset~C) prevent sharp transient instability but excite slow-moving, system-wide oscillatory modes visible in the energy flow metrics. This indicates that even when the system maintains frequency stability, it can still experience considerable stress that manifests in the directional energy flow patterns, making these metrics valuable indicators for comprehensive stability assessment.


\subsection{System-wide Small-signal Impacts: Snapshot-based Assessments}

The small-signal stability of the grid is sensitive to the sharp ramping behavior of LDDLs. Under traditional operating conditions, system operators typically monitor the system at a fixed operating point and evaluate the damping of dominant modes, such as inter-area oscillations. Since load and generation vary gradually in conventional grids, these eigenvalue-based estimates remain valid over extended time horizons, offering a dependable representation of system dynamics and stability margins. In contrast, rapid increases or decreases in LDDL demand can induce abrupt shifts in operating points, causing eigenvalue trajectories to move quickly. Such movements may result in temporary reductions in damping or even the appearance of poorly damped oscillatory modes, which can remain undetected if stability is assessed only at a few isolated operating points.

To overcome this limitation, a snapshot-based analysis framework is required. By evaluating the system at multiple points along the LDDL ramping trajectory and computing the corresponding eigenvalue spectra, operators can track the evolution of critical modes and their damping characteristics in real time. This approach makes it possible to identify regions where stability margins are most vulnerable and to gain deeper insight into the underlying mechanisms, such as converter control dynamics, network interactions, or protection responses, that influence these variations. The enhanced visibility provided by snapshot-based methods supports proactive mitigation, including adjustments to ramping schedules, fine-tuning of controller parameters, or the deployment of stabilizing resources such as damping controllers and energy storage systems, thereby improving grid resilience under fast-changing load conditions.

Let us consider the grid's state variables as $x \in \mathbb{R}^n$ including LDDL interface states, algebraic variables $p \in \mathbb{R}^r$, control inputs $u\in \mathbb{R}^m$, and LDDL input variables $v \in \mathbb{R}^s$ such as load consumptions, thereby, the dynamical equation can be compactly written as:
\begin{align}
    \dot{x} = f(x,u,p,v),
\end{align}
and the power flow as:
\begin{align}
    g(x,p,u,v)= 0,
\end{align}
where $f(.)$ denotes the nonlinear dynamics, and $g(.)$ denotes the nonlinear power flows. Linearizing along the initial operating point, we get,
\begin{align}
    \Delta \dot{x} = A_{11}\Delta x + A_{12}\Delta u + A_{13}\Delta p + A_{14}\Delta v,\\
    0 = A_{21}\Delta x + A_{22}\Delta u + A_{23}\Delta p + A_{24}\Delta v,
\end{align}
We have,
\begin{align}
    \Delta p = -A_{23}^{-1}A_{21}\Delta x - A_{23}^{-1}A_{22}\Delta u - A_{23}^{-1}A_{24}\Delta v
\end{align}
Therefore, the dynamical model can be captured as:
\begin{align}
    \Delta \dot{x} &= (A_{11}- A_{13}A_{23}^{-1}A_{21})\Delta x + (A_{12} - A_{13}A_{23}^{-1}A_{21})\Delta u \nonumber \\  &+(A_{14} - A_{13}A_{23}^{-1}A_{21})\Delta v,
\end{align}

We can consider without any additional supplementary control action as:
\begin{align}
    \Delta \dot{x} = (A_{11}- A_{13}A_{23}^{-1}A_{21})\Delta x  +  (A_{14} - A_{13}A_{23}^{-1}A_{21})\Delta v,
\end{align}
The Jacobians in the state matrix are denoted as:
\begin{align}
    A_{11} = \frac{\partial f}{\partial x}|_{\alpha_0}, A_{13} = \frac{\partial f}{\partial p}|_{\alpha_0}, A_{23} = \frac{\partial g}{\partial p}|_{\alpha_0}, A_{21} = \frac{\partial g}{\partial x}|_{\alpha_0}.
\end{align}
The state matrix at that operating point $\alpha_0$ is denoted as: $A_0 = (A_{11}- A_{13}A_{23}^{-1}A_{21})$.
As the data center loads vary over time during a ramping behavior, the operating conditions are also varied from $\alpha_0$ to $\alpha_1, \alpha_2,\dots,\alpha_k$ resulting in the following sequence of state matrices for one particular ramping condition within a small time window dictated by $\tilde{k}$ as the upper bound on the time indices:
\begin{align}
    \mathcal{A} = \{ A_0, A_1, \dots, A_k \}, k \leq \tilde{k}.
\end{align}
As these state matrices are functions of the operating conditions, which are in turn functions of the LDDL load variables, we can capture the changes as:
\begin{align}
    \mathcal{A} = \{ A_0(v_0), A_1(v_1), \dots, A_k(v_k) \}, k \leq \tilde{k}.
\end{align}
Therefore, the LDDL consumption impacts the dynamics in two ways: the forcing term in the dynamical equation, and the perturbation in the state matrices due to the perturbation in the operating conditions. From the operating condition $\alpha_i$ to $\alpha_{i+1}$, the perturbation in the state matrices caused by the LDDL consumption change $\delta v: (v_{i+1} - v_{i})$ is given as:
\begin{align}
    \Delta A_i = \frac{\partial A_i}{\partial v_i} \delta v, A_{i+1} = A_{i} + \Delta A_i
\end{align}


\noindent \textit{Proposition 1:} Considering a compact set of LDDL consumption ramp $V = \{v_0, v_1, \dots, v_k\}$ converging to $v^*$ with state matrix $A^*$, the perturbation in the eigenvalues of the state matrix is given by, 
\begin{align}
    \Delta \lambda_i = \frac{w_i^* (\frac{\partial A_i}{\partial v_i} \delta v) v_i}{w_i^* v_i},
\end{align}
and such a sequence converges within the compact set of eigenvalues to a limiting stable load condition $v^*$ of a particular ramp. \qed

The proposition follows from recalling the convergence of spectra in terms of the Hausdorff metric distance. Let $(X,d)$ be a metric space, and let $A,B \subset X$ be nonempty compact sets.  
The \emph{Hausdorff distance} \cite{kraft2020computing, gil2021new} between $A$ and $B$ is defined by
\[
d_H(A,B) = \max \left\{
    \sup_{a \in A} \inf_{b \in B} d(a,b), \;
    \sup_{b \in B} \inf_{a \in A} d(a,b)
\right\}.
\] 
For this analysis, the underlying metric space is the complex plane $\mathbb{C}$, or more specifically, nonempty complex subsets of $\mathbb{C}$ with the distance metric as $ d = |\gamma_1 - \gamma_2|,$ where $\gamma_1, \gamma_2 \in \mathbb{C}$. We have the sequence $(A_k)$ with spectra $\sigma(A_k)$,  
and $A^*$ with spectrum $\sigma(A^*)$, which will have $
\sigma(A_k) \to \sigma(A^*) \; \text{in the Hausdorff sense} $
with
$
d_H\big( \sigma(A_k), \sigma(A^*) \big) \!\longrightarrow\! 0.
$

However, this only considers the set convergence, and the system may be subjected to bifurcations; therefore, the safe set will be a subset of the compact convergent set. 

\noindent \textit{Proposition 2:} Suppose that the spectra converge in the Hausdorff sense:
$
\sigma(A_n) \xrightarrow{d_H} \Sigma.
$ The \emph{safe set} of eigenvalues $S$ is then a subset of $\Sigma$, 
defined by system constraints such as
$
\Re(\lambda) < 0.
$ Under bifurcations, $\Sigma$ may contain eigenvalues 
outside the safe region, even though Hausdorff convergence holds, i.e,
$
S \subseteq \Sigma. $

Alg. \ref{alg:stability_analysis} describes the snapshot-based small-signal stability metric-based monitoring algorithm.

\begin{algorithm}[h!]
\SetAlgoLined
\caption{Snapshot-Based Small-Signal Stability Analysis under LDDL Ramp}
\label{alg:stability_analysis}
\DontPrintSemicolon
{\bf Inputs:} Dynamic model $f(.)$, ramp profile $v(t)$, thresholds $\zeta_{\min}, m_{\min}$.\;
{\bf Outputs:} Stability margins, critical modes, visualization data.\;

Define snapshot set $\{v(t_k)\}_{k=1}^N$ along ramp.\;

\For{$k = 1, \ldots, N$}{
    Solve steady-state equilibrium: $x^*(t_k) \gets$ power flow.\;
    Linearize system: $A_k \gets \partial f/\partial x \big|_{(x^*(t_k),v(t_k))}$.\;
    Compute eigenvalues $\{\lambda_i^k\}$ of $A_k$.\;
    Evaluate metrics:\;
    \Indp Spectral abscissa: $m_k = -\max_i \Re(\lambda_i^k)$.\;
    Damping ratio: $\zeta_{k,\min} = \min_i \frac{-\Re(\lambda_i^k)}{\sqrt{(\Re(\lambda_i^k))^2+(\Im(\lambda_i^k))^2}}$.\;
    \Indm Identify critical mode $\lambda^*_k$ achieving $\zeta_{k,\min}$.\;
    Compute participation factors for $\lambda^*_k$.\;

    \If{$m_k \leq m_{\min}$ \textbf{or} $\zeta_{k,\min} \leq \zeta_{\min}$}{
        Flag snapshot $v(t_k)$ as critical.\;
    }
}

\end{algorithm}

\begin{figure*}[t]
    \centering
    \subfigure[]{\label{fig:damping_ratio_dataset_A}\includegraphics[width=0.329\linewidth]{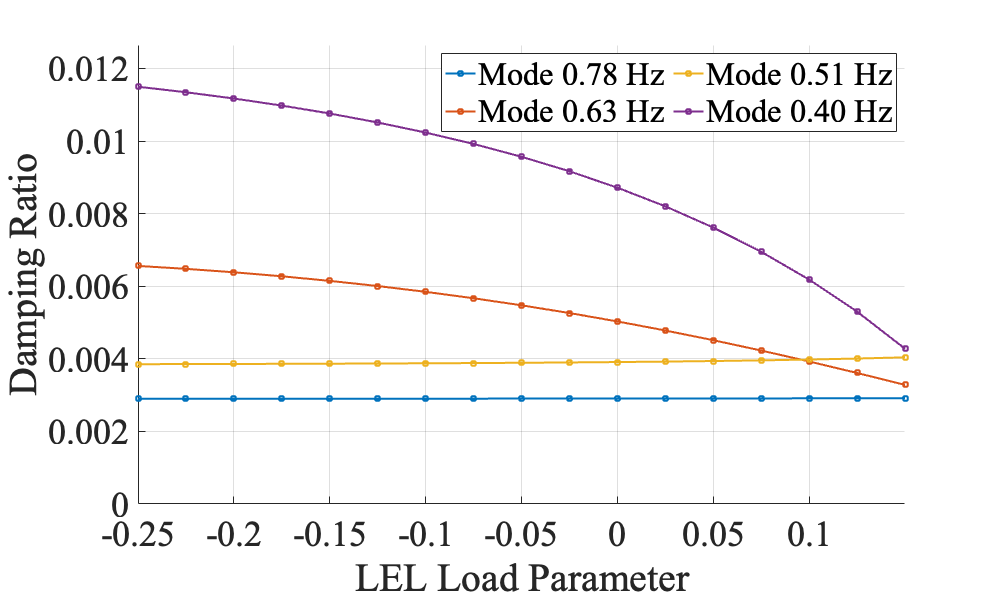}}
    \subfigure[]{\label{fig:damping_ratio_dataset_B}\includegraphics[width=0.329\linewidth]{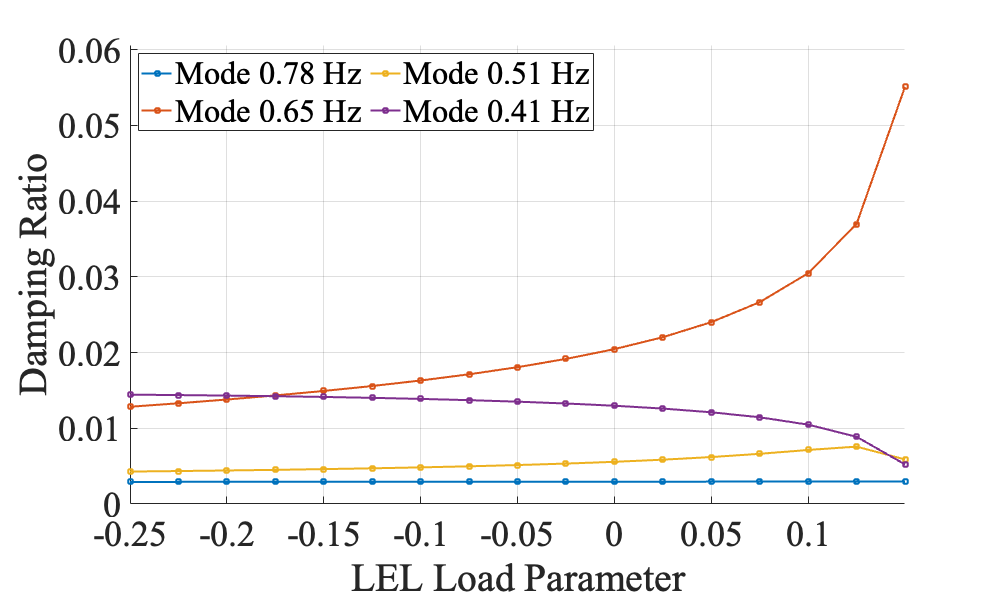}}
    \subfigure[]{\label{fig:damping_ratio_dataset_C}\includegraphics[width=0.329\linewidth]{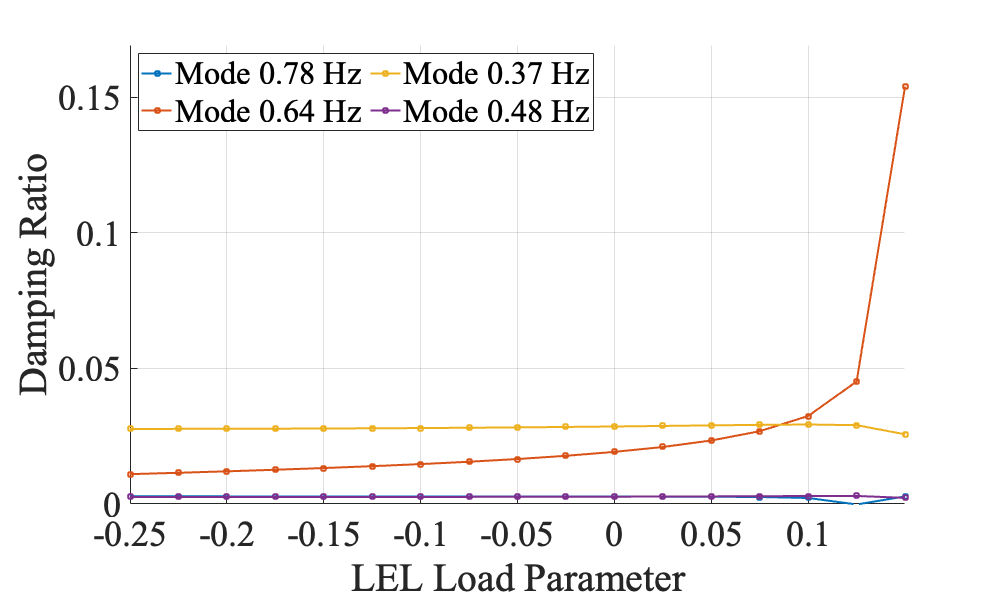}}
    \subfigure[]{\label{fig:realpart_dataset_A}\includegraphics[width=0.329\linewidth]{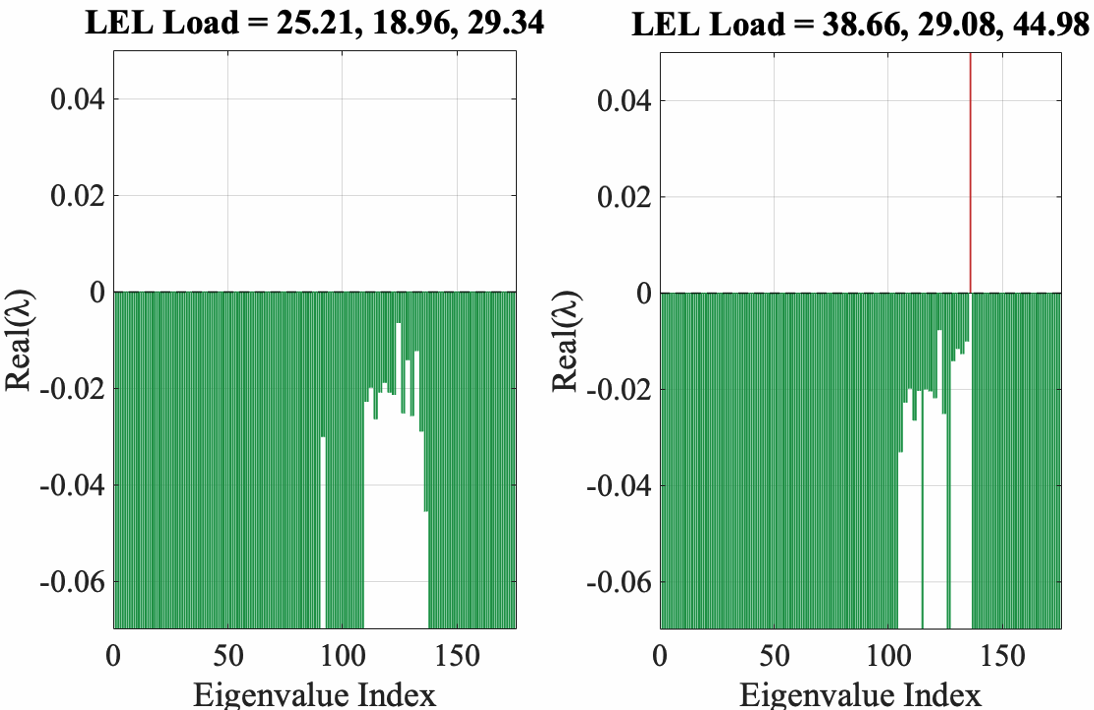}}
    \subfigure[]{\label{fig:realpart_dataset_B}\includegraphics[width=0.329\linewidth]{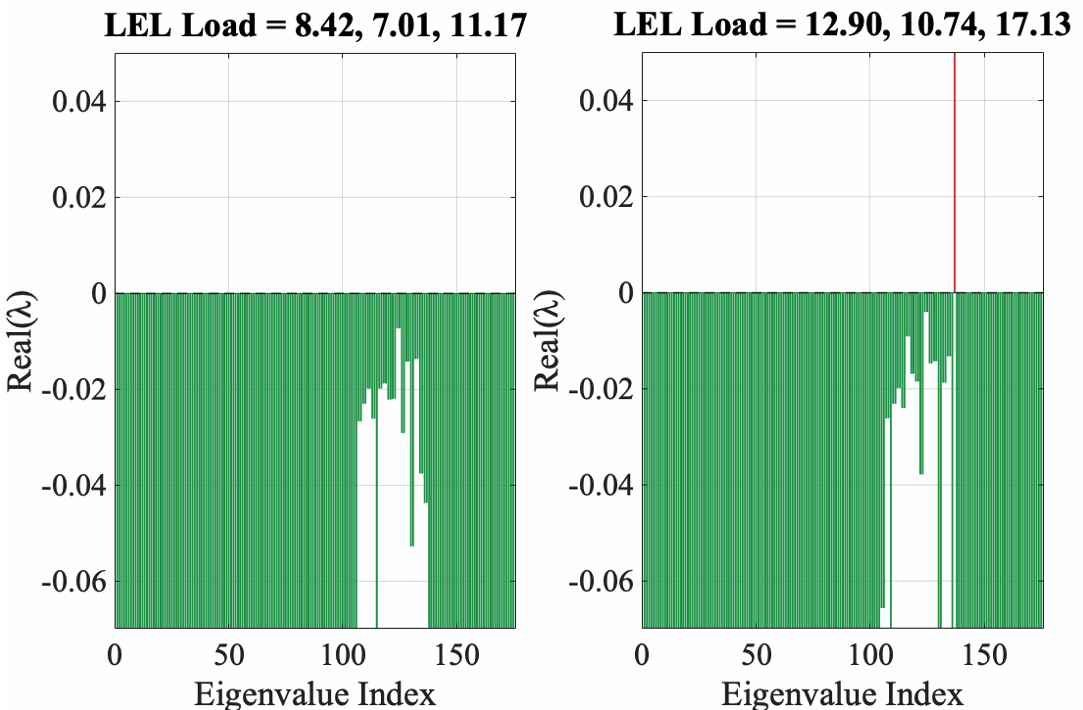}}
    \subfigure[]{\label{fig:realpart_dataset_C}\includegraphics[width=0.329\linewidth]{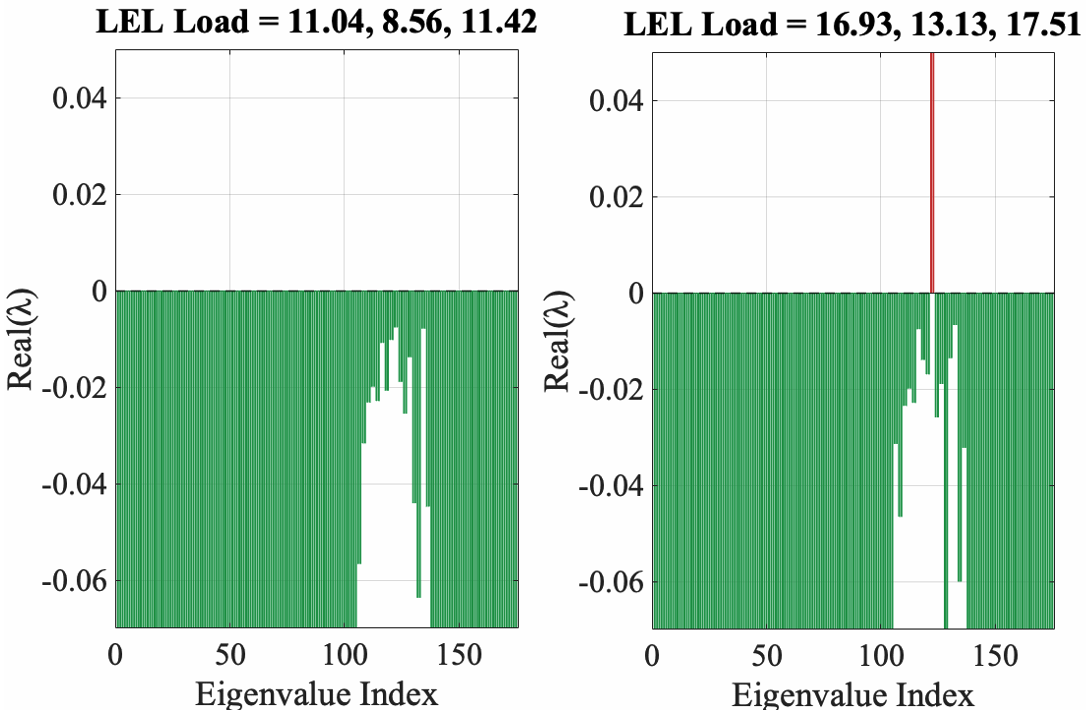}}
    \caption{Small-signal Analysis result: damping ratio trajectory of (a) Scenario~A, (b) Scenario~B, (c) Scenario~C and the eigenvalue realpart example ($-25$\% and $+15$\%) of (d) Scenario~A, (e) Scenario~B, (f) Scenario~C}
    \label{fig:ssm_result}
    \vspace{-0.3cm}
\end{figure*}

\vspace{-0.2 cm}
\subsection{Numerical Experiments with Small-Signal Studies}

To demonstrate the efficacy of the snapshot-based small-signal analysis framework outlined in Alg.~\ref{alg:stability_analysis}, we conduct numerical experiments on a modified IEEE 68-bus test system. We investigate three different scenarios for the placement of LDDLs, designated as scenarios A, B, and C. For each scenario, we vary the LDDL load from $-25$\% (representing a load decrease) to $+15$\% (representing a load increase) of its nominal value and analyze the trajectory of the system's dominant inter-area modes.

\textit{Scenario A:} LDDLs are located at buses $9, 36$, and $37$. The results of the analysis are presented in Fig.~\ref{fig:damping_ratio_dataset_A} and Fig.~\ref{fig:realpart_dataset_A}. First, Fig.~\ref{fig:damping_ratio_dataset_A} reveals a critical trend: as the LDDL load increases, the damping ratios of the 0.63 Hz and 0.40 Hz modes progressively decrease. The damping of the $0.40$ Hz mode, in particular, deteriorates from approximately $1.15$\% to below $0.5$\%, a level considered critically low for secure grid operation. This degradation culminates in instability, as confirmed by Fig.~\ref{fig:realpart_dataset_A}. The system remains stable at a $-25$\% load decrease. In contrast, at a $+15$\% load increase, an eigenvalue crosses into the right-half plane (highlighted in red), rendering the system unstable. Participation factor analysis shows that the unstable mode is impacted by the inverter angles at buses $12$ and $25$, showing complex interactions caused by the integrated data centers with existing grid components.

\textit{Scenario B:} LDDLs are placed at buses $15$, $16$, and $20$ in Area~1 as indicated in Fig.~\ref{fig:system}. To compute a feasible power flow solution with a nominal value of LDDL, we reduce the load of these three buses to $80$\% of the load represented in Fig.~\ref{fig:lel_load}, such as $17.95, 14.95$ and $23.83$ p.u.
The analysis for this scenario illustrates the location-dependent nature of LDDL impacts. Unlike scenario A, Fig.~\ref{fig:damping_ratio_dataset_B} shows that increasing the LDDL load has a mixed effect on damping. The damping of the $0.65$ Hz mode significantly improves, rising from around $1.2$\% to over $5$\%. However, the damping of the $0.41$ Hz mode simultaneously degrades. This opposing behavior highlights a key challenge: a change that is beneficial for one mode may be detrimental to another. As seen in Fig.~\ref{fig:realpart_dataset_B}, despite the improved damping of one mode, an eigenvalue associated with a different mode becomes unstable at the $+15$\% load point. Here also, participation factor analysis shows that the inverter angles at the boundary of area $1$ and $2$ at buses $1, 3$, and $8$ cause this bifurcation.

\textit{Scenario C:} LDDLs are connected at buses $3, 4$, and $18$, in the middle of Area~1  (referring to Fig.~\ref{fig:system}). Similar to Scenario~B, we reduce the nominal loads to $70$\% from the values in Fig.~\ref{fig:lel_load} ($16.48, 12.79, 17.05$ p.u.). This case provides another distinct result. Here, an increase in LDDL load leads to a dramatic improvement in the damping of the $0.64$ Hz mode, as its damping ratio dramatically increases from $1.5$\% to nearly $16$\% in Fig.~\ref{fig:damping_ratio_dataset_C}. This suggests that, at certain locations, LDDLs can make the system stable for specific inter-area oscillations. However, a closer look at the full spectrum in Fig.~\ref{fig:realpart_dataset_C} reveals that even in this scenario, a different, previously well-damped mode is driven to instability by the increased loading. This shows the critical importance of the snapshot-based framework; focusing only on the most prominent or historically problematic modes can obscure emerging threats from other parts of the eigenvalue spectrum. Similar to the previous scenarios, participation factor analysis reveals that the angles of the data center at bus $3$ and the inverter at bus $12$ have the most impact on the unstable mode, demonstrating how LDDL integration at different locations can shift the sources of system instability across various network components.

\section{Conclusion} \label{sec:CON}
This paper investigated the stability implications of large dynamic digital loads (LDDLs), with a particular focus on AI-driven data centers. For nonlinear transient behavior, we introduced energy-flow-based metrics that capture both localized and coupling stress at data center buses. The results showed that abrupt spikes create severe frequency deviations, sustained oscillations can escalate into collapse, and gradual ramps excite slower oscillatory modes. These insights highlight how the proposed energy-flow analytics provide a fine-grained view of transient stress that conventional stability measures fail to reveal. For small-signal behavior, we developed a snapshot-based analysis framework that tracks eigenvalue trajectories during rapid load ramps. This approach revealed how different modes can simultaneously improve or deteriorate in damping, depending on location and loading conditions, showing the importance of continuous monitoring rather than single-point assessments. The main contribution of this work lies in advancing stability-aware assessment tools that bridge transient and small-signal domains, offering operators deeper situational awareness of evolving risks from data center integration. Future work will extend these methods by incorporating detailed dynamics of LDDLs and coordinated storage, and by validating the proposed metrics with real-time measurements to enable their integration into operator-facing decision-support platforms.

\section*{Acknowledgment}

The research is supported by the Energy and Environment Directorate's Laboratory Directed Research at Pacific Northwest National Laboratory (PNNL). The authors would like to thank Soumya Kundu, Sai Pushpak Nandanoori, Ramij R. Hossain, and Bowen Huang at Pacific Northwest National Laboratory for sharing LDDL consumption profile examples and helpful discussions related to this work. Authors would also like to thank Long Vu and Brett Ross at Pacific Northwest National Laboratory for helpful suggestions and discussions on the LDDL research. 

\bibliographystyle{IEEEtran}
\bibliography{ref}

\appendices

\section{Additional Modeling Details:}

We consider a bulk power system which includes synchronous generators (SGs), grid-forming inverters (GFMs), and  large dynamic digital loads (LDDLs). We utilize the IEEE $68-$bus benchmark models in phasor domain with the network parameters are obtained from the standard data set \cite{ieee68} with additional modifications to interconnect inverters and LDDLs.

The dynamics of each SG are governed by the classical swing equations that are sufficient for power oscillation related stability studies and frequency dynamics~\cite{kundur2007power}:
\begin{subequations}
\label{eq:sg}
\begin{align}
\dot{\delta}_i &= \omega_i - \omega_0,\\
\dot{\omega}_i &= \tfrac{1}{M_i}\!\left[D_i(\omega_0 - \omega_i) + P_i - P_{ei}\right],
\end{align}
\end{subequations}
where $\delta_i$ and $\omega_i$ denote the rotor angle and frequency of generator $i$, $P_i$ is the mechanical input power, and $P_{ei}$ is the electrical output power. The constants $M_i$ and $D_i$ represent the inertia and damping coefficients, respectively.  The IEEE $68-$bus test system is modified to include GFM IBRs on select buses. There are a total of 35 such inverters each at a load bus with few of them only contains the LDDLs (to create the AI data center hub). Each LDDL at bus $j \in \mathcal{L}$ is modeled as a grid-interactive load whose power consumption is modulated by an interfacing inverter which mimics installations of interactive UPS, storage or local generation to support the LDDL. The control structure for the interfacing inverter is adapted from droop-based principles to regulate power exchange with the grid, more specifically  REGFM A1 model \cite{du2023model} developed at PNNL. The corresponding dynamic equations are:
\begin{subequations}
\label{eq:LDDL}
\begin{align}
\dot{\delta}_j &= \omega_j - \omega_0, \\
\dot{\omega}_j &= \tfrac{1}{\tau_j}\!\left[\omega_0 - \omega_j + m_{p_j}(P_j^{set} -{P}^{L}_{j}- P_j)\right], \label{eq:LDDL_omega} \\
\dot{V}^e_j &= \tfrac{1}{\tau_j}\!\left[V_j^{set} - V_j - V^e_j + m_{q_j}(Q_j^{set} - Q_j)\right], \\
\dot{E}_j &= k^{pv}_j \dot{V}^e_j + k^{iv}_j V^e_j, \\
\dot{P}^{L}_{j} &= \tfrac{1}{T_{Lj}}(P^{AI}_j - {P}^{L}_{j}). \label{eq:LDDL_filter}
\end{align}
\end{subequations}
Here, $\delta_j$ and $\omega_j$ are the voltage angle and frequency of the inverter interface. The variables $V_j$ and $E_j$ denote the terminal and internal voltage magnitudes of the inverter, and $V^e_j$ is an auxiliary state for the voltage controller. The parameters $m_{p_j}$ and $m_{q_j}$ are the \Pfdroop\ and \QVdroop\ droop coefficients, while $k^{pv}_j$ and $k^{iv}_j$ are the proportional and integral gains of the voltage control loop. The setpoints $P_j^{set}$, $Q_j^{set}$, and $V_j^{set}$ are the desired active power consumption, reactive power exchange, and terminal voltage magnitude, respectively. The variables $P_j$ and $Q_j$ are the active and reactive power consumed by the other local loads at the inverter terminals and grid if the inverter interfaces a local generation, and the LDDL consumption is captured as $P^{AI}_j$.  In equation \eqref{eq:LDDL_filter}, $P^{L}_{j}$ is the intermediate state representing the power consumed by the internal power electronics of the load. This first-order filter, with time constant $T_{Lj}$, models the aggregate dynamic properties of the interfacing power electronics, such as the DC power distribution unit (and fast dynamics of rectifier). It effectively mimics the energy buffering (dynamic filtering) effect of interfacing elements on the $P^{AI}_j$.
The active and reactive power balance at each bus $j = 1, \dots, N$ is expressed as:
\begin{subequations}
\label{eqn:load_flows}
\begin{align}
0 &= P_{ej} - \mathrm{Re}\!\left\{\sum_{k=1,\,k \ne j}^{N} V_j (V_{jk} B_{jk})^*\right\} - V_j^2 G_j, \label{eqn:load_active}\\
0 &= Q_{ej} - \mathrm{Im}\!\left\{\sum_{k=1,\,k \ne j}^{N} V_j (V_{jk} B_{jk})^*\right\} - V_j^2 B_j, \label{eqn:load_reactive}
\end{align}
\end{subequations}
where $P_{ej}$ and $Q_{ej}$ are the net active and reactive power injections at bus $j$. For a bus $j$ with an LDDL, these injections are negative consumptions, i.e., $P_{ej} = -P_j$ and $Q_{ej} = -Q_j$. The terms $G_j$ and $B_j$ denote the shunt conductance and susceptance at bus $j$, and $B_{jk}$ is the susceptance of the lossless line between buses $j$ and $k$. In compact form, the network equations are:
\begin{equation}
0 = g(x_s, x_l, V),
\end{equation}
where $x_s$ and $x_l$ are the state vectors for the SGs and LDDLs, respectively, and $V$ is the vector of bus voltage magnitudes.

\section{Additional Details on the Directional Energy Flow Numericals} 

\begin{figure*}[t!]
    \centering
    \subfigure[]{\label{fig:directional_coupling_energy_flow_data_A}\includegraphics[width=0.32\linewidth]{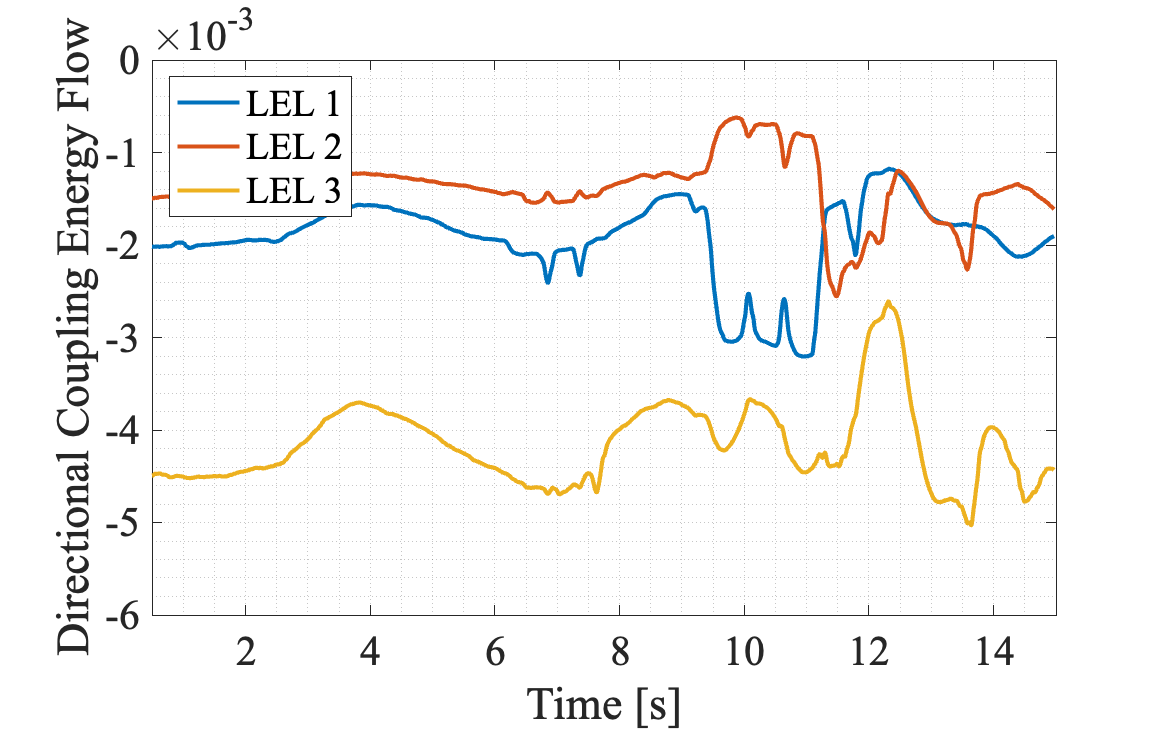}}
    \subfigure[]{\label{fig:neighboring_directional_energy_flow_data_A}\includegraphics[width=0.32\linewidth]{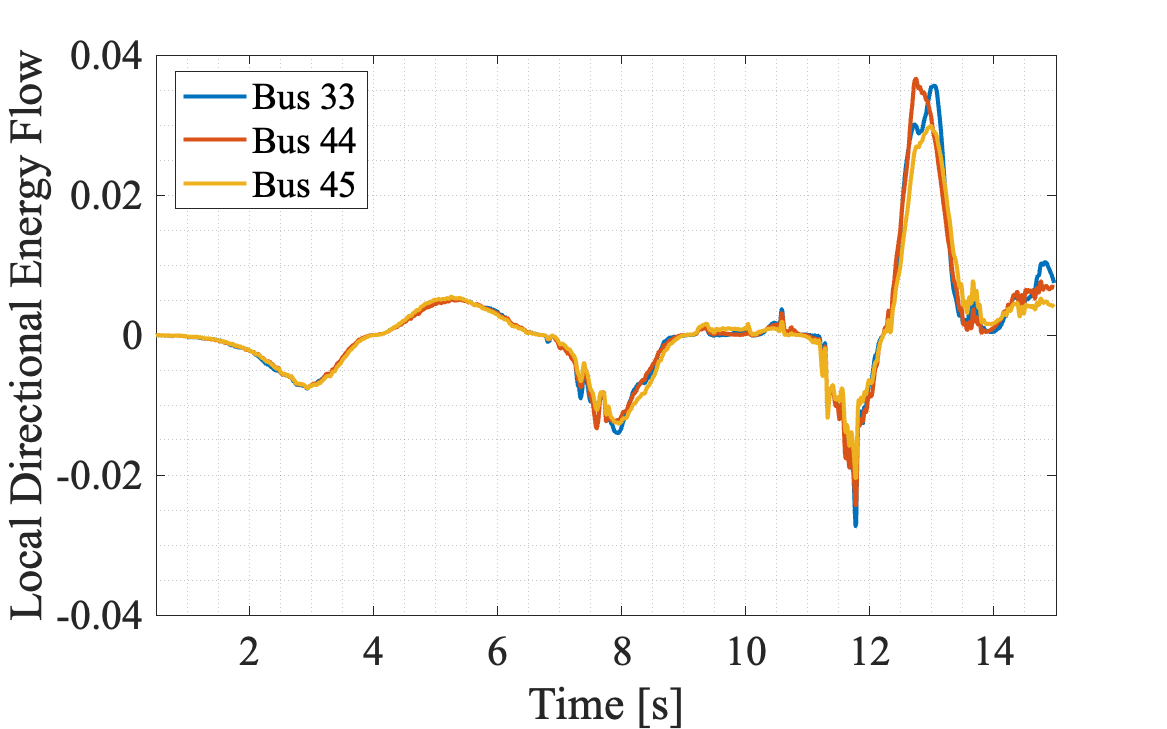}}
    \subfigure[]{\label{fig:rocof_data_A}\includegraphics[width=0.32\linewidth]{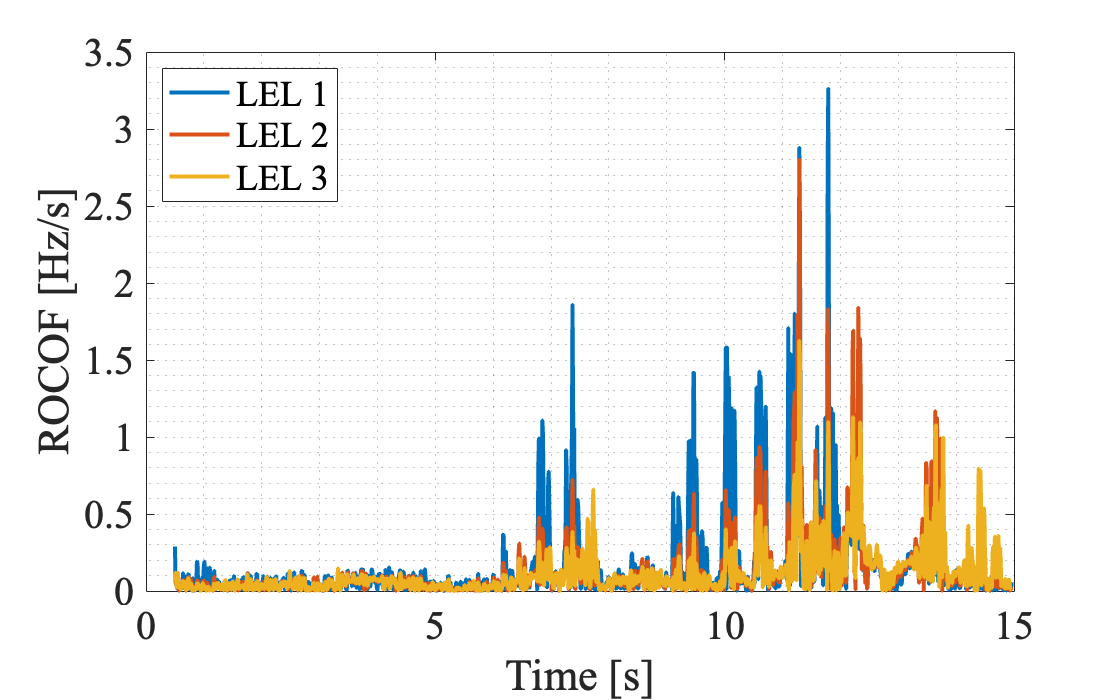}}
    \subfigure[]{\label{fig:directional_coupling_energy_flow_data_B}\includegraphics[width=0.32\linewidth]{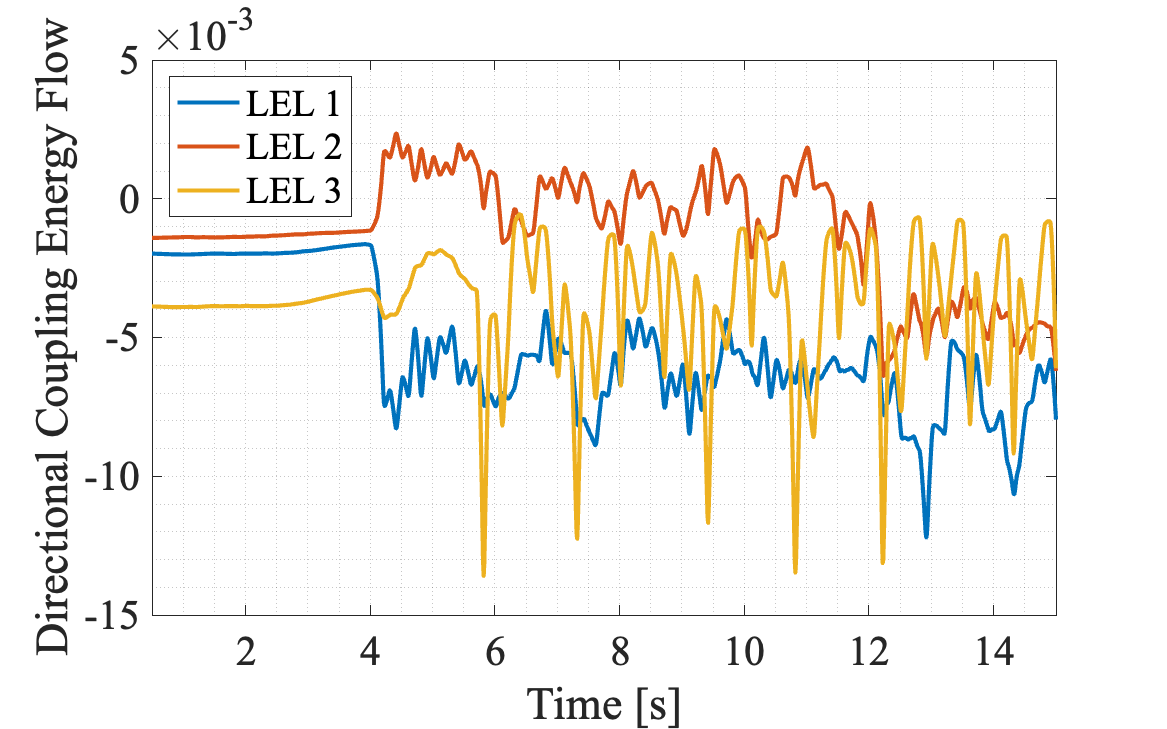}}
    \subfigure[]{\label{fig:neighboring_directional_energy_flow_data_B}\includegraphics[width=0.32\linewidth]{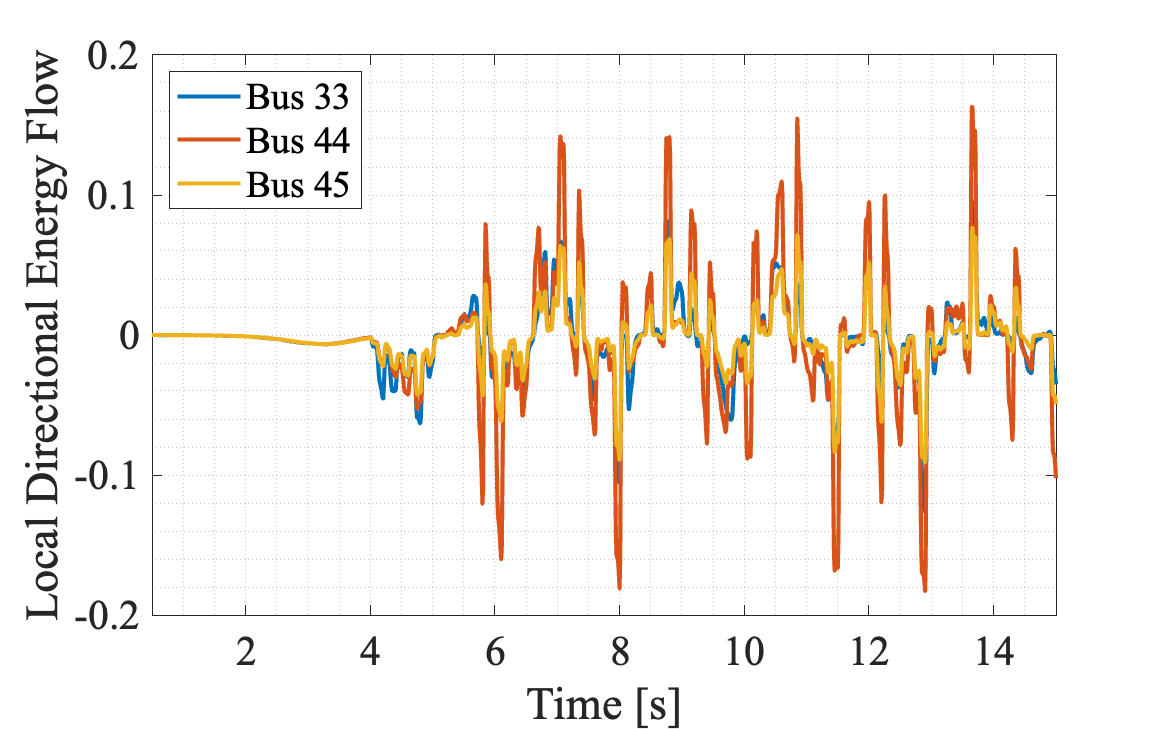}}
    \subfigure[]{\label{fig:rocof_data_B}\includegraphics[width=0.32\linewidth]{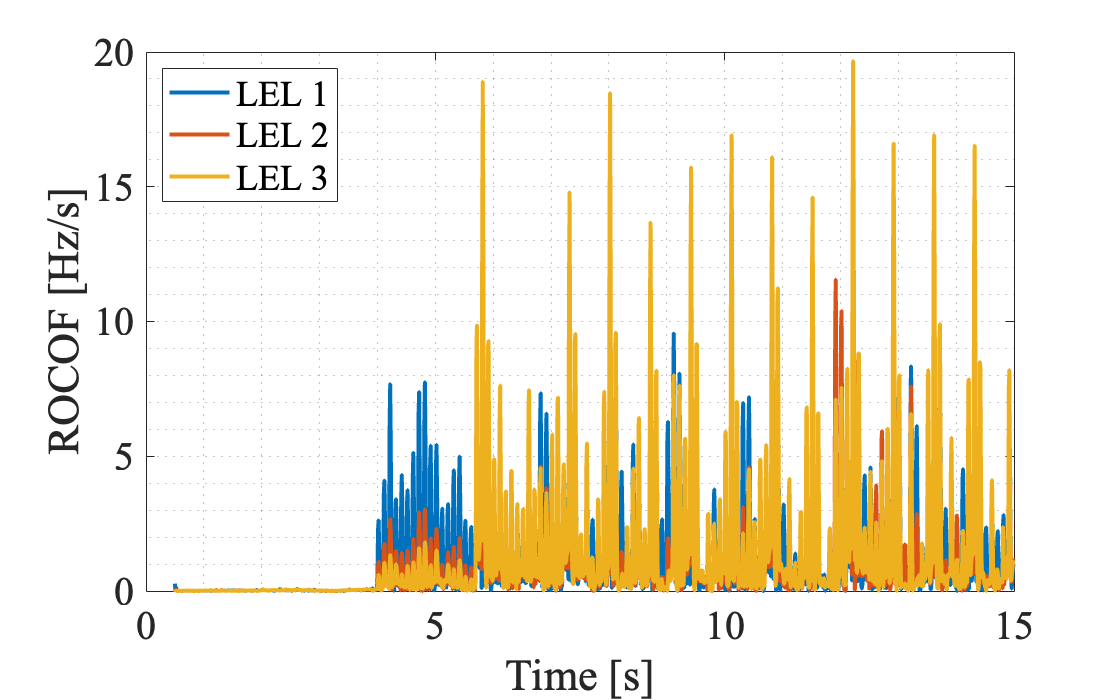}}
    \subfigure[]{\label{fig:directional_coupling_energy_flow_data_C}\includegraphics[width=0.32\linewidth]{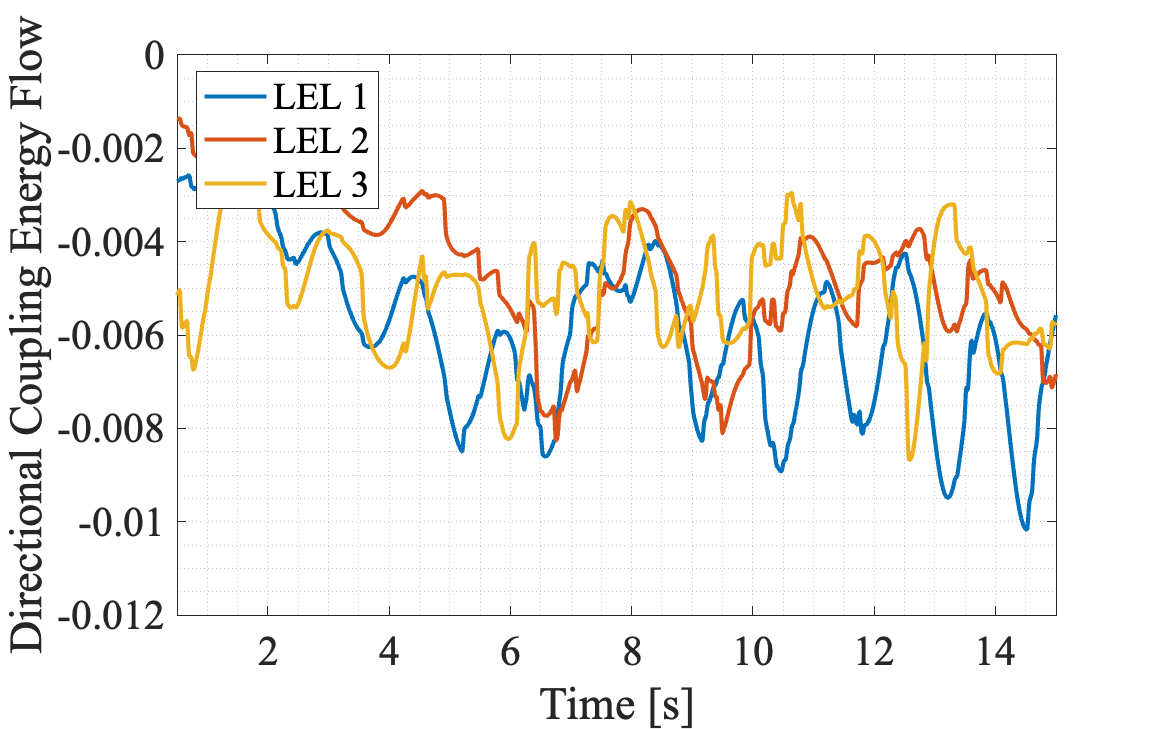}}
    \subfigure[]{\label{fig:neighboring_directional_energy_flow_data_C}\includegraphics[width=0.32\linewidth]{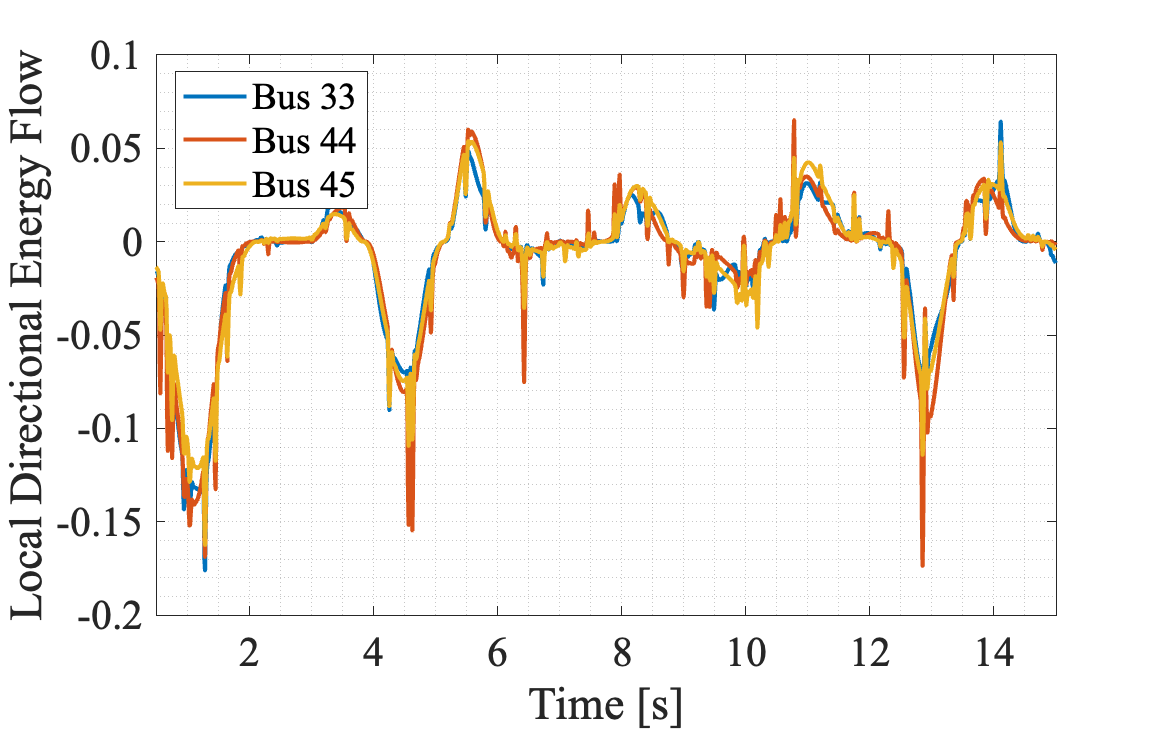}}
    \subfigure[]{\label{fig:rocof_data_C}\includegraphics[width=0.32\linewidth]{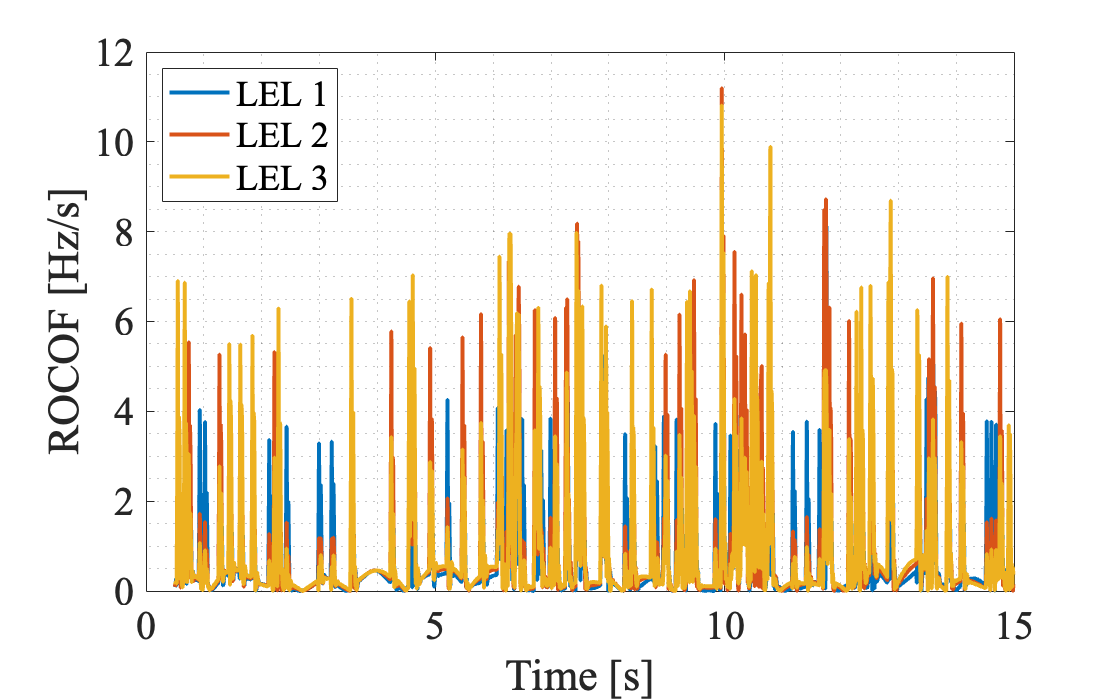}}
    \caption{Directional coupling and local energy flow characteristics under Scenarios A, B, and C. 
    Subplots show (a)--(c) directional coupling energy flow, (d)--(f) local directional energy flow of neighboring nodes near the data center bus, and (g)--(i) rate of change of frequency (RoCoF). 
    Scenarios A, B, and C correspond to distinct data center operational patterns characterized by sharp inference spikes, sustained oscillatory consumption, and gradual HPC-style ramp-up behaviors, respectively.}
    \label{fig:supp_directional_flows}
\end{figure*}

Fig.~\ref{fig:supp_directional_flows} illustrates the directional coupling energy flow, local directional energy flow at neighboring buses, and the corresponding rate of change of frequency (RoCoF) across three representative operational scenarios. 
Panels (a)--(c) correspond to Scenario~A, panels (d)--(f) correspond to Scenario~B, and panels (g)--(i) correspond to Scenario~C. 
Each scenario represents a distinct load behavior at the data center: (A) sharp inference-type spikes, (B) sustained oscillatory consumption, and (C) gradual high-performance computing (HPC)-style ramping. 
The plots provide a comparative view of how these operational differences manifest in the dynamic energy flow patterns and frequency responses.

In \textbf{Scenario~A} [Figs.~\ref{fig:supp_directional_flows}(a)--(c)], the data center undergoes rapid and high-magnitude fluctuations that cause pronounced variations in both the directional coupling energy and local directional energy flow. 
These quantities exhibit sharp peaks and fast oscillations, reflecting strong transient interactions between the data center bus and its neighboring nodes. 
The RoCoF trace also shows abrupt changes aligned with these energy flow spikes, indicating that the system experiences short but intense frequency disturbances following each load spike. 
The strong temporal alignment among the three metrics highlights the impulsive nature of the load events in this scenario.

In \textbf{Scenario~B} [Figs.~\ref{fig:supp_directional_flows}(d)--(f)], the system response becomes dominated by sustained oscillatory components. 
The directional coupling energy exhibits slower but persistent oscillations of larger amplitude compared to Scenario~A, suggesting prolonged energy exchanges between interconnected buses. 
The local directional energy flow at the neighboring buses follows a similar trend but remains at smaller magnitudes, indicating that nearby inverters and local buses experience comparable yet attenuated responses. 
The RoCoF waveform shows quasi-periodic oscillations that persist throughout the time window, corresponding to the sustained load fluctuations characteristic of this scenario. 
Overall, Scenario~B represents a condition where continuous oscillatory stress rather than impulsive events governs the system dynamics.

In \textbf{Scenario~C} [Figs.~\ref{fig:supp_directional_flows}(g)--(i)], the gradual ramp-up in data center load results in much smoother temporal profiles for both coupling and local directional energy flows. 
The variations are moderate and evolve over longer time scales, indicating that the system adapts to the changing load with minimal abrupt transients. 
The local directional energy flow at the neighboring nodes again follows the same overall pattern as that of the data center bus, but with smaller magnitudes. 
The RoCoF trace shows small and well-damped oscillations, reflecting a stable frequency response under gradual load variations. 
Compared with the previous two cases, Scenario~C demonstrates that slowly varying loads cause less severe transient stress, producing smoother and more regular energy flow trajectories.

Taken together, these results show that the temporal characteristics of the data center load directly influence the shape and intensity of both directional and local energy flows, as well as their corresponding frequency responses. 
While Scenario~A is marked by fast impulsive interactions, Scenario~B exhibits persistent oscillations, and Scenario~C shows gradual, well-damped dynamics. 
The similar temporal patterns but smaller amplitudes in the neighboring nodes further indicate that local inverter-connected buses experience comparable but attenuated directional energy flow behaviors relative to the main data center bus.

\section{Participation Factor Analysis Details}

The participation factors \cite{ieee68} utilized in Section~III.D are computed using both right and left eigenvectors to identify which state variables contribute most significantly to each eigenvalue (mode). For a power system with state matrix $\mathbf{A}$, the participation factor $p_{ik}$ of the $i$-th state variable in the $k$-th mode is mathematically defined as:

\begin{equation}
p_{ik} = \frac{|\phi_{ik} \psi_{ik}|}{\sum_{j=1}^{n} |\phi_{jk} \psi_{jk}|}
\end{equation}

where $\phi_{ik}$ and $\psi_{ik}$ represent the $i$-th elements of the $k$-th right and left eigenvectors, respectively, and $n$ is the total number of state variables \cite{kundur2007power}. The right eigenvectors $\boldsymbol{\phi}_k$ satisfy $\mathbf{A}\boldsymbol{\phi}_k = \lambda_k \boldsymbol{\phi}_k$, while the left eigenvectors $\boldsymbol{\psi}_k$ satisfy $\boldsymbol{\psi}_k^T \mathbf{A} = \lambda_k \boldsymbol{\psi}_k^T$.

The computational procedure involves element-wise multiplication of the right eigenvector with the complex conjugate of the corresponding left eigenvector, followed by taking the absolute value:

\begin{equation}
\mathbf{P} = |\boldsymbol{\Phi} \odot \overline{\boldsymbol{\Psi}}|
\end{equation}

where $\boldsymbol{\Phi}$ and $\boldsymbol{\Psi}$ are matrices containing right and left eigenvectors as columns, $\odot$ denotes element-wise multiplication, and $\overline{\cdot}$ represents complex conjugation. Each column of the resulting participation factor matrix is then normalized such that the sum of participation factors for each mode equals unity. This normalization ensures that participation factors can be interpreted as the relative contribution of each state variable to the corresponding eigenmode, facilitating the identification of dominant system components in modal behavior.

The participation factor analysis reveals the dominant state variables contributing to the most unstable eigenmode across three different scenarios. Tables~\ref{tab:pf_scenario_a}, \ref{tab:pf_scenario_b}, and \ref{tab:pf_scenario_c} present the top five participating state variables for the most critical unstable mode in each scenario, focusing on the rotor angle states (Delta) that exhibit the highest participation factors.

\begin{table}[t!]
\centering
\caption{Participation factors for the most unstable mode in Scenario A}
\label{tab:pf_scenario_a}
\begin{tabular}{ccccc}
\toprule
\textbf{Eigenvalue Real Part} & \textbf{Component} & \textbf{State} & \textbf{Bus} & \textbf{PF} \\
\midrule
0.090124 & Inverter & Delta & 12 & 0.487 \\
0.090124 & Inverter & Delta & 25 & 0.336 \\
0.090124 & Inverter & Delta & 18 & 0.038 \\
0.090124 & Inverter & Delta & 21 & 0.021 \\
0.090124 & Data center & Delta & 9 & 0.020 \\
\bottomrule
\end{tabular}
\end{table}

\begin{table}[t!]
\centering
\caption{Participation factors for the most unstable mode in Scenario B}
\label{tab:pf_scenario_b}
\begin{tabular}{ccccc}
\toprule
\textbf{Eigenvalue Real Part} & \textbf{Component} & \textbf{State} & \textbf{Bus} & \textbf{PF} \\
\midrule
0.065186 & Inverter & Delta & 3 & 0.406 \\
0.065186 & Inverter & Delta & 1 & 0.280 \\
0.065186 & Inverter & Delta & 8 & 0.195 \\
0.065186 & Inverter & Delta & 27 & 0.036 \\
0.065186 & Data center & Delta & 20 & 0.018 \\
\bottomrule
\end{tabular}
\end{table}

\begin{table}[t!]
\centering
\caption{Participation factors for the most unstable mode in Scenario C}
\label{tab:pf_scenario_c}
\begin{tabular}{ccccc}
\toprule
\textbf{Eigenvalue Real Part} & \textbf{Component} & \textbf{State} & \textbf{Bus} & \textbf{PF} \\
\midrule
0.109327 & Data center & Delta & 3 & 0.393 \\
0.109327 & Inverter & Delta & 12 & 0.379 \\
0.109327 & Inverter & Delta & 16 & 0.055 \\
0.109327 & Inverter & Delta & 26 & 0.052 \\
0.109327 & Inverter & Delta & 23 & 0.022 \\
\bottomrule
\end{tabular}
\end{table}

The results demonstrate that rotor angle deviations (Delta) of inverters and data centers are the primary contributors to system instability across all scenarios. In Scenarios A and B, inverter-based resources dominate the participation factors, with buses 12 and 25 in Scenario A, and buses 3 and 1 in Scenario B showing the highest contributions, showing complex interactions caused by the integrated data centers with existing grid components. For Scenario B, the inverter angles at the boundary of area $1$ and $2$ at buses $1, 3$, and $8$ cause this bifurcation. Notably, Scenario C exhibits a more balanced participation between data centers and inverters, with the data center at bus 3 having the highest participation factor of 0.393, closely followed by the inverter at bus 12 with 0.379. This analysis provides critical insights for targeted control strategies and system reinforcement planning.

\end{document}